\documentclass[a4paper,11pt]{article}
\pdfoutput=1 

\usepackage{jheppub} 
\usepackage[T1]{fontenc} 
\usepackage[italian,english]{babel}
\usepackage{hyperref}
\usepackage{ifpdf}
\usepackage{subfigure}
\usepackage{tikz}
\usepackage[compat=1.1.0]{tikz-feynman}
\usetikzlibrary{arrows,shapes}
\usetikzlibrary{trees}
\usetikzlibrary{matrix,arrows} 				
\usetikzlibrary{positioning}				
\usetikzlibrary{calc,through}				
\usetikzlibrary{decorations.pathreplacing}  
\usepackage{pgffor}							

\usetikzlibrary{decorations.pathmorphing}	
\usetikzlibrary{decorations.markings}
\tikzset{
	vector/.style={decorate, decoration={snake}, draw},
	fermion/.style={draw=black, postaction={decorate}}, 
	scalar/.style={dashed,draw=black, postaction={decorate}}}
\tikzstyle{block} = [draw, rectangle, 
minimum height=3em, minimum width=6em]

\usepackage{amssymb}
\usepackage{amsfonts}
\usepackage{epsf}
\usepackage{rotating}
\usepackage{graphicx}
\usepackage{amsmath}
\usepackage{fancyhdr}
\usepackage{lineno}
\usepackage{babel}
\usepackage{graphics}
\usepackage{pstricks}
\usepackage{color}
\usepackage{multirow}
\usepackage{float}

\def\sn#1{\textit{Scenario-#1}}
\newcommand{\nn}{\nonumber}
\newcommand{\lsim}{\mathrel{\mathop{\kern 0pt \rlap
			{\raise.2ex\hbox{$<$}}}
		\lower.9ex\hbox{\kern-.190em $\sim$}}}
\newcommand{\gsim}{\mathrel{\mathop{\kern 0pt \rlap
			{\raise.2ex\hbox{$>$}}}
		\lower.9ex\hbox{\kern-.190em $\sim$}}}

\newcommand{\be}{\begin{equation}}
\newcommand{\ee}{\end{equation}}
\newcommand{\bea}{\begin{eqnarray}}
\newcommand{\eea}{\end{eqnarray}}

\def\sn#1{\textit{Scenario-#1}}

\def\gev{\ensuremath{\mathrm{\,Ge\kern -0.1em V\,}}}
\def\tev{\ensuremath{\mathrm{\,Te\kern -0.1em V\,}}}

\newcommand{\MM}{\textcolor{black}}
\newcommand{\MMc}{\textcolor{black}}

\title{\boldmath  Relativistic Freeze-in with Scalar Dark Matter in a Gauged $B-L$ Model   and Electroweak Symmetry Breaking}

\author[a]{Priyotosh Bandyopadhyay,}
\author[b,c]{Manimala Mitra,}
\author[b,c]{Abhishek Roy}

\newcommand{\AddrHBNI}{
	Homi Bhabha National Institute, BARC Training School Complex, Anushakti Nagar, Mumbai 400094, India }
	
\affiliation[a]{Indian Institute of Technology Hyderabad, Kandi,  Sangareddy-50228, Telengana, India}
\affiliation[b]{Institute of Physics, Sachivalaya Marg, Bhubaneswar, Pin-751005, Odisha, India}
\affiliation[c]{\AddrHBNI}

\emailAdd{bpriyo@phy.iith.ac.in} 
\emailAdd{manimala@iopb.res.in} 
\emailAdd{abhishek.r@iopb.res.in }

\preprint{IITH-PH-0009/20 \\ \hspace*{0pt}\hfill IP/BBSR/2020-7}

\abstract{We explore  relativistic freeze-in production of scalar dark matter in gauged $B-L$ model, where we focus on the production of dark matter from the decay and annihilation of Standard Model (SM) and $B-L$ Higgs bosons. We consider the Bose-Einstein (BE) and Fermi-Dirac (FD) statistics,  along with the thermal mass correction of the SM Higgs boson in our analysis. We show that in addition to the SM Higgs boson, the annihilation and decay of the $B-L$ scalar can also contribute substantially to the dark matter relic density.  Potential effects of electroweak symmetry breaking (EWSB) and  thermal mass correction in BE framework enhance the dark matter relic substantially as it freezes-in near EWSB temperature via scalar annihilation. However, such effects are not so prominent when the dark matter freezes-in at a later epoch than EWSB, dominantly by decay of scalars. The results of this analysis are rather generic, and applicable to  other similar scenarios.}

\begin{document}
\maketitle
\flushbottom
	\section{Introduction}
	
The null-results from a number of dark matter direct detection experiments motivate to explore alternate dark matter production mechanisms. One of the most well-motivated dark matter production mechanisms is  freeze-in \cite{Hall:2009bx} production of dark matter. In this framework, the dark matter is feebly coupled  with the Standard Model (SM) particles, in general with particles in equilibrium and  thereby  referred as feebly interacting massive particle (FIMP). Due to very suppressed interaction, the FIMP dark matter never attains thermal equilibrium with particles with which they are feebly coupled. The suppressed interaction further gives natural explanation for the non-observation  of any  direct detection signal. The dark matter in freeze-in scenario is produced from the decay and/or annihilation of SM and beyond Standard Model (BSM) particles which are either in equilibrium \cite{Molinaro:2014lfa, Biswas:2015sva, Merle:2015oja, Shakya:2015xnx, Konig:2016dzg, Biswas:2016iyh, Biswas:2016yjr, Biswas:2017tce} or also freezing-in along side the dark matter \cite{Bandyopadhyay:2020qpn}. We explore freeze-in  production of dark matter in extended gauged $B-L$ model, where we address few of the subtlety of the production.  

The gauged $B-L$ model \cite{Mohapatra:1980qe, Wetterich:1981bx, Georgi:1981pg} is one of the most appealing, yet minimal theory descriptions, that explain small SM neutrino masses. The model includes three right handed neutrinos (RH neutrinos) required  for anomaly cancellation, one $B-L$ gauge boson, and a complex scalar field. The scalar field acquires vacuum expectation value, and breaks the $B-L$ gauge symmetry. The $B-L$ gauge boson, as well as, the heavy neutrinos acquire their masses due to $B-L$ symmetry breaking. The light neutrinos, on the other hand, acquire their masses via seesaw \cite{Mohapatra:1979ia, minkowski1977mu}, with their masses inversely proportional to the $B-L$ symmetry breaking scale. A scalar particle with $B-L$ charge can be accommodated in this model which serves as the dark matter candidate by suitable choice of $B-L$ charge. The freeze-out scenario for this model has been explored in  \cite{ Sanchez-Vega:2014rka, Guo:2015lxa, Singirala:2017see, Klasen:2016qux, Okada:2010wd, Rodejohann:2015lca}, along with other phenomenological implications. The late decay of RH neutrinos are explored in scalar dark matter  freeze-out scenario \cite{Bandyopadhyay:2017bgh, Bandyopadhyay:2018qcv}.  In this context the fermionic dark matter has also been explored \cite{Basak:2013cga, Okada:2016gsh, Okada:2016tci, Kaneta:2016vkq, Okada:2012sg, Abdallah:2019svm, Biswas:2016bfo}. The freeze-in scenario along with  neutrino mass and leptogenesis  has been studied for this model in \cite{Biswas:2017tce}.  The freeze-out scenario via semi-annihilation has been explored in \cite{Rodejohann:2015lca}. 

One of the most crucial parameter, the $B-L$ charge of the scalar dark matter, {\it i.e.,} $q_{DM}$ is not guided by the model, rather is a free parameter. The non-observation of any direct detection signal motivates to choose a very small value of the charge $q_{DM}$. Furthermore, a choice of very small $q_{DM}$ suppresses interactions with other particles in  equilibrium,   leaving out the freeze-out framework completely. In this context some studies have been pursued \cite{Biswas:2017tce,Biswas:2016bfo, Kaneta:2016vkq,Abdallah:2019svm}, where $B-L$ gauge boson still contributes in the freeze-in production of the dark matter. The fermion plus scalar dark matter freeze-in scenarios are also explored \cite{Chianese:2019epo, Chianese:2018dsz, Chianese:2020khl}.

In this article we explore the regime, where the $B-L$ gauge boson contribution is negligible in dark matter relic density, and the freeze-in dynamics is governed by annihilation and decays of SM and $B-L$ scalars. To evaluate the relic density, we adopt the relativistic framework \cite{Lebedev:2019ton,DeRomeri:2020wng, Arcadi:2019oxh}, where we use Bose-Einstien (BE), and Fermi-Dirac (FD) statistics. The effect of thermal mass correction of SM Higgs boson along with electroweak symmetry breaking (EWSB)  on dark matter phenomenology has been explored in \cite{Baker:2017zwx,Heeba:2018wtf} for singlet scalar extension. We explore such effects of EWSB, thermal mass correction, and quantum statistics for scalar extended gauged $B-L$ model.  A number of SM and BSM decay and annihilation modes become open at different epoch of the early Universe, which we carefully include in our numerical computation. Similar to \cite{DeRomeri:2020wng}, which found large enhancement in fusion process, we find significant  enhancement in $2 \to 2$ annihilation, and $1 \to 2$ decay  processes during EWSB, once thermal mass correction of SM Higgs boson has been taken into account. Though our study is confined to gauge $B-L$ framework, the results of this analysis are more generic for mainly two reasons: i) the freeze-in here is dominant by the scalar, and not so by the $B-L$ gauge boson, ii) the relativistic effects at EWSB that we observe are applicable to more generic scenario. The gauged $B-L$ model has also been explored for  collider phenomenologies. In the $B-L$ scenario, the RH neutrinos are charged under $B-L$ gauge group and thus they can be produced at the colliders via $Z_{BL}$ unlike the Type I seesaw case  \cite{Bandyopadhyay:2017bgh, Deppisch:2018eth, Banerjee:2015hoa, Deppisch:2019ldi, Deppisch:2019kvs}. 

Depending on the primary production mechanisms,  we classify a few different scenarios \sn{1-5}. In {\it Scenario-1,\,2} the freeze-in of dark matter production is controlled by the annihilation of the SM and $B-L$ Higgs  boson. 
For  {\it Scenario-3,\,4,\,5},  it is rather dominated by the decays of SM and $B-L$ scalar. For numerical analysis, we include  all other annihilation processes, involving   other SM particles, and RH neutrinos. We present a comparison of the relic density, obtained using BE distributions and MB distributions. We
 observe,  that for annihilation dominant freeze-in scenario, if freeze-in occurs at  EWSB, the relic density using BE distribution is larger as compared to MB distribution. For scenarios, where  freeze-in occurs at a  later epoch than EWSB, the enhancement is relatively suppressed, as the reaction rates using BE statistics and MB statistics become very similar during the freeze-in epoch.
Overall, we find that thermal mass correction of SM Higgs boson and EWSB have a large impact on the production of the dark matter in this model, that can most accurately be described by quantum statistics.  

The paper is organised as follows. In Section.~\ref{model}, we describe the model. Following this in Section.~\ref{freezeinprod}, we discuss the dark matter production in various scenarios. We present our conclusion in Section.~\ref{conclu}. In Appendix.~\ref{appen2}, and \ref{appen1},  we provide the necessary calculation details. 
	
\section{The Model}\label{model}	
\MM{We consider gauged $B-L$ model that contains one  SM gauge singlet complex scalar field $\mathcal{S}$ and three heavy right handed neutrinos (RH-neutrinos) $N_i$. In this theory framework,  the vacuum expectation value (vev) of the gauge singlet scalar field breaks the $B-L$ symmetry.  Additionally, we also consider another SM gauge singlet complex scalar field $\phi_D$, which we consider to be the   dark matter.} The $\Phi$ and $L$ are the SM Higgs and  $SU(2)_L$ lepton doublets. Other than the scalar  fields $\phi_D$, and $\mathcal{S}$, the   $N_i (i=1,2,3)$ states are also singlet under  SM gauge group \cite{Bandyopadhyay:2017bgh,Rodejohann:2015lca, Biswas:2017tce}. 
The Majorana masses are generated by the spontaneous breaking of the $B-L$ symmetry. \MM{We show  the charge assignments of different multiplets  in Table~\ref{tabbml}. The dark matter $\phi_{D}$ is non-trivially charged with a charge $q_{DM}$ under $U(1)_{B-L}$.}
\MM{The RH-neutrinos interact with the SM lepton doublet, SM Higgs field and the complex scalar field $\mathcal{S}$ through  Yukawa couplings $y_{N}^{\prime},$ and $\lambda_{NS}$,  as shown in Eq.~\eqref{lag}}. The scalar potential of the model with $\Phi$, $\mathcal{S}$ and $\phi_D$ fields contains few additional terms, as compared to the SM. The Yukawa Lagrangian involving $\mathcal{S}$, $N_i$ and $\phi_{D }$ fields, and the scalar potential are given by,  
\bea
\mathcal{L}_\text{BSM}&=&  - \mu_S^2 |\mathcal{S}|^2- \mu_h^2 |\Phi|^2-  \mu_D^2 |\phi_D|^2 -\lambda_{Sh} {|\mathcal{S}|}^2 |\Phi|^2  -\lambda_{ SD}|\phi_{D}|^2|\mathcal{S}|^2 -\lambda_{D h}|\phi_{D}|^2|\Phi|^2 \nn\\
&&-\lambda_{h} |\Phi|^4-\lambda_{S} |\mathcal{S}|^4 -\lambda_{D}|\phi_{D}|^4 \nn \\  &&-\left(\sum_{i=1}^{3} \lambda_{NS}  {\mathcal{S}} \bar{N}_{i}^{c} N_{i} + \sum_{i, j=1}^{3} y_{N,i j}^{\prime} \bar{L}_{i} \tilde{\Phi} N_{j}+h.c.\right) .
\label{lag}
\eea
As is evident from the above Lagrangian, the model contains quartic interactions involving dark matter-Higgs, as well as dark matter-$\mathcal{S}$ fields, that have major impact in determining the dark matter relic abundance.  

\begin{table}[h]
	\centering
	\begin{tabular}{ c c c c cc cc cc}
		\hline 
		& $\Phi$ & $N$ &  $L$ & $Q$ & $u_R$ & $d_R$ & $e_R$ & $\mathcal{S}$&$\phi_{DM}$\\ \hline
		$Y_{B-L}$ & $0$ & $-1$ & $-1$ & $1/3$ &$1/3$ & $1/3$ &$1$& $2$&$q_{DM}$ \\
		\hline
	\end{tabular}
	\caption{$B-L$ charges for all the fields present in the model.}  \label{tabbml}
\end{table}
Other than these particles, the model also contains $B-L$ gauge boson $Z_{BL}$. See \cite{Mohapatra:1980qe, Wetterich:1981bx, Georgi:1981pg} for detail descriptions of the model. Below, we present a brief dicussion on neutrino masses, the scalar and gauge sector of the model, which would be relevant for our subsequent analysis.
\begin{itemize}
\item {\it Gauge boson mass:\ }The additional gauge boson from $U(1)_{B-L}$ is represented by $Z_{BL}$ where the mass of $Z_{BL}$ is generated due to spontaneous  breaking of the $B-L$ gauge symmetry, and is given by, 
\be
m_{Z_{\rm BL}}=2g_{BL}v_{BL}.
\ee
In the above $g_{BL}$ represents $B-L$ gauge coupling and  the vev of $\mathcal{S}$ is denoted by $v_{BL}$. The $\sqrt{s}=13$ TeV LHC search for a massive resonance decaying into di-lepton final states puts a strong lower bound on the $Z_{B-L}$ gauge boson mass, {\it i.e., } $m_{Z_{B-L}} > 5.15$ TeV \cite{CMS:2019tbu}. For our calculation, we consider $m_{Z_{B-L}}=5.5$ TeV, which is in agreement with the LHC bound. 

\item {\it Scalar masses:\ }Owing to the non-zero $\lambda_{Sh}$ term in Eq.~\ref{lag}, and non-zero vev's $v, v_{BL} $,  the scalar fields $\mathcal{S}$ and $\Phi$ mix with each other after electroweak symmetry breaking. We define the neutral components of $\mathcal{S}$ and $\Phi$ fields as $S+iS_I$ and $h+ih_I$,  respectively, which leads to the mass matrix of $h$ and $S$ after EWSB as, 
\begin{equation}
\mathcal{M}_{scalar}^{2}=\begin{pmatrix}
2\lambda_{h}v^2 & \lambda_{Sh} v_{BL}v\\
\lambda_{Sh} v_{BL}v & 2\lambda_{S}v_{BL}^2
\end{pmatrix}.
\end{equation}
Rotating the basis $h$ and $S$ to new states $h_1$ and $h_2$ by suitable angle $\alpha$, we can diagonalise the above mass matrix. The physical mass basis are given by,
\begin{equation}
\begin{split} 
h_1 &=  h\cos{\alpha}+S\sin{\alpha}, \\ 
h_2 &=  -h\sin{\alpha}+S\cos{\alpha},
\end{split}
\end{equation}
where $h_1$ is the SM-like Higgs boson and $h_2$ is the BSM scalar.  The mixing angle between them is given by,
\begin{equation}
\tan{2\alpha}=\frac{\lambda_{Sh}vv_{BL}}{\lambda_{h}v^{2}-\lambda_{S}v_{BL}^2}. 
\end{equation}
The mass square eigenvalues of scalar field $h_1$ and $h_2$ are given by, 
\begin{equation}
m_{h1,h2}^2=\lambda_h v^2+\lambda_{ S}v_{BL}^2\mp \sqrt{(\lambda_{h} v^2-\lambda_{S}v_{BL}^2)+(\lambda_{ Sh}vv_{BL})^2}. 
\label{eq:massesscalar}
\end{equation}

For our analysis we stay in the decoupling limit {\it i.e.,} $\alpha \sim 10^{-4} - 10^{-5}$ obeying $2\sigma$ constraints of Higgs data of LHC at 13 TeV \cite{Sirunyan:2018koj,ATLAS:2018doi}.
\MM{Therefore, for all practical purposes, due to the very tiny mixing between the SM Higgs and $B-L$ Higgs bosons  $h_1 \simeq h$ and $h_2 \simeq S$ in our analysis. In the subsequent sections, we explore  the production of dark matter from the SM and $B-L$ Higgs boson decay and annihilation processes. For the above mentioned values of the  Higgs mixing angle $\alpha$ and quartic coupling $\lambda_{Sh} > 6 \times 10^{-6}$ \cite{DeRomeri:2020wng},  the $B-L$ Higgs boson is in thermal equilibrium along with SM Higgs boson in the early Universe. } 

Note that Eq.~\eqref{eq:massesscalar} represents the physical masses of the scalar fields without any  thermal correction and we will see that the thermal  correction to the SM Higgs mass have a large  impact on the dark matter phenomenology. The electroweak phase transition (EWSB) can be either second order phase transition or a cross-over. The SM Higgs becomes massless in second order phase transition, whereas it remains massive in cross-over during EWSB \cite{Baker:2017zwx,Quiros:1999jp,DOnofrio:2015gop}. The authors have performed numerical lattice Monte Carlo simulations to study 
the thermodynamics of the cross-over where they have shown that $m_h (T_c)$ approaches around $10 - 15\ GeV$ during EWSB \cite{DOnofrio:2015gop}. In our work,  we consider the electroweak phase transition to be a crossover in which Higgs remains massive at critical temperature ($T_c=160\, \textrm{GeV}$).  For the temperature is greater than the critical temperature  {\it i.e.,} $T>T_c$, the mass of Higgs bosons is given by \cite{DeRomeri:2020wng}, 
\begin{equation}\label{mhT}
m_{h}^{2}(T)= c(T^2-T_{c}^2 ) + m_{h}^{2}(T_c),
\end{equation}
whereas for the temperature smaller than the critical temperature  {\it i.e.,} $T <T_c$, the mass of Higgs boson is given by, 
\begin{equation}
m_{h}^{2}(T)=2 c(T_{c}^2-T^2 ) + m_{h}^{2}(T_c).
\end{equation}
In the above,  $c$ represents  a constant which is determined by requiring $m_{h}(0)=125.5\, \textrm{GeV}$, {\it i.e.,} Higgs boson mass at zero temperature.
	 \begin{figure}[h]
	\begin{center} 
		\includegraphics[width=0.7\linewidth,angle=-0]{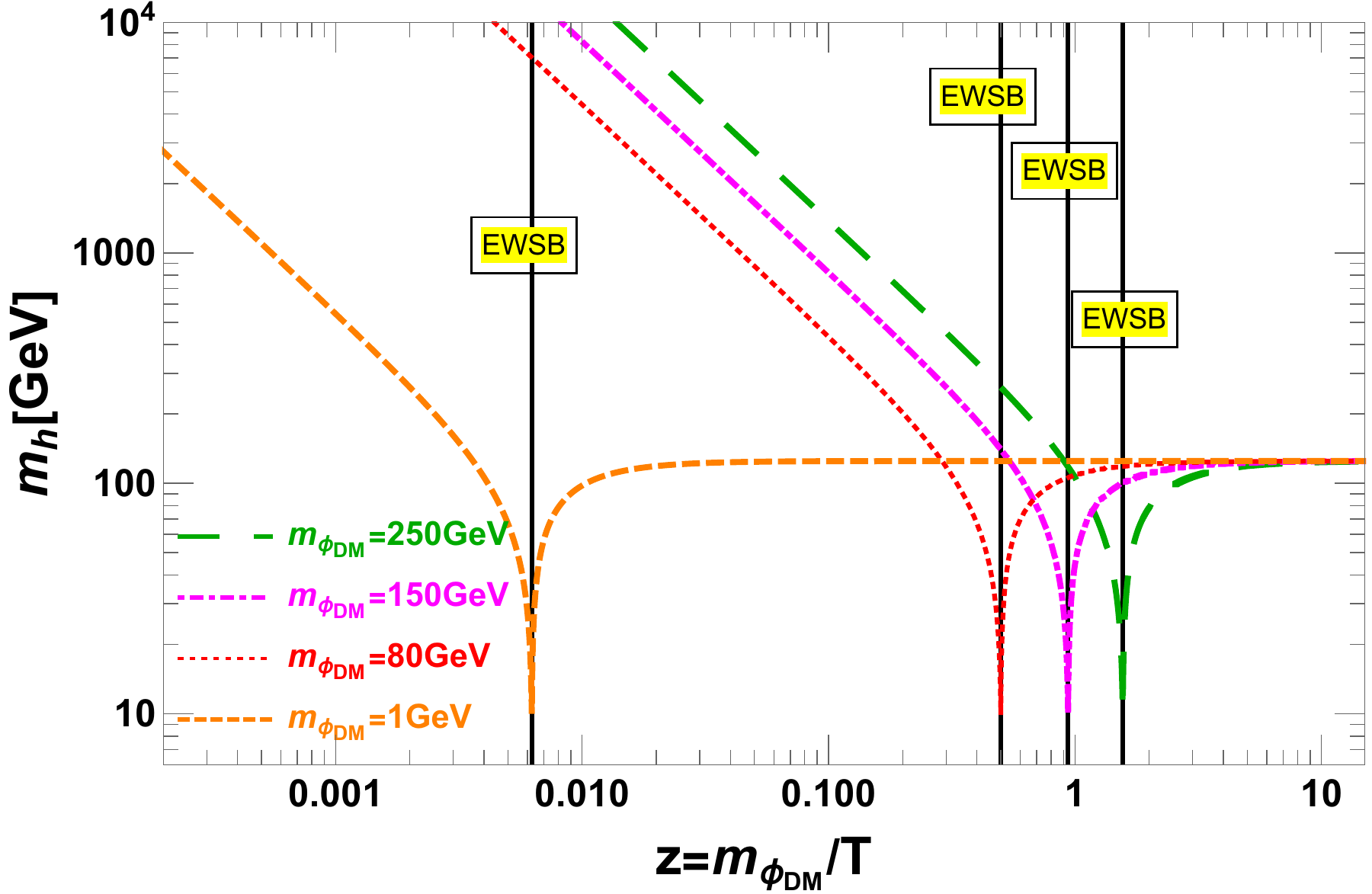}
		\caption{Variation of the SM Higgs boson mass with $z$ for different scenarios, {\it Scenario-1} to {\it Scenario-5}. The different scenarios correspond to  $m_{\phi_{DM}}=250,150,80$ GeV for {\it Scenario-1,2,3}, and $m_{\phi_{DM}}=1$ GeV for {\it Scenario-4,5}. \label{fmh}}
	\end{center}
\end{figure}
In Fig.~\ref{fmh} we show  thermal corrections to the SM Higgs boson mass for different scenarios which we detail later. The vertical lines represent the $z=\frac{m_{\phi_{DM}}}{T}$ values corresponding to  EWSB and for our analysis $m_h(T_c)\approx 10\, \textrm{GeV}$.
Note that, due to the difference in the DM mass,  EWSB ($T_{EW}= 160\ GeV$) corresponds to different values of $z$ for these different scenarios. This is clearly evident   from Fig.~\ref{fmh}.  EWSB has a significant importance in our work, as it will be clear from the discussions of the subsequent sections.\newline
Similarly,  the thermal correction for the mass of $B-L$ scalar $S$ can also be calculated. However, for our analysis this is not so important and it can be understood easily as follows. To evaluate thermal correction, the parameter $\mu_S$ in the Lagrangian Eq.~\eqref{lag} should be replaced by \cite{DeRomeri:2020wng}, 
\begin{equation}
\mu_S^{2}\to \mu_S^{2}+c_S T^2,
\end{equation}
where 
\begin{equation}
c_S \approx \frac{1}{4}\lambda_{ S}+\frac{1}{6}\lambda_{ Sh}. 
\end{equation}
The critical temperature $T_c^{v_{BL}}$ is the temperature, where $B-L$ scalar $\mathcal{S}$ takes vaccuum expectation value $v_{BL}$ and breaks the $U(1)_{B-L}$ symmetry. The critical temperature $T_c^{v_{BL}}$ then can be approximated as \cite{DeRomeri:2020wng}, 
\begin{equation}
 T_c^{v_{BL}} \approx \frac{|\mu_S|}{\sqrt{c_S}}.
 \label{reheat}
\end{equation}

In our present work, we consider that the $B-L$ breaking took place at a high temperature $T_c^{v_{BL}}$ in the early Universe, which we assume to be equal to the re-heating temperature of the Universe \cite{Fukuyama:2005tf, Borah:2020wyc}. Using Eq.~\eqref{reheat} for our parameter choices, we  obtain that  the re-heating temperature to be $T_R \sim 2.26\times10^{4}\  \textrm{GeV}$. Immediately after the re-heating or $B-L$ symmetry breaking, the field $S$ acquires a mass $200$ GeV, that we consider throughout our analysis. Hence, thermal correction to $S$ mass is not relevant in our study.
\item {\it Neutrino masses:\ }The masses of the  light SM neutrinos are  generated via  the usual Type-I seesaw mechanism:	
\be
{m^{\nu}_{ij}}=\frac{y_{N,i k}^{\prime} y_{N,k j}^{\prime}<\Phi^2>}{m_{N, k}},
\ee
where $m_{N,k}= \lambda_{NS}  \langle\mathcal{S}\rangle $ are the  Majorana masses of the RH neutrinos generated due to the spontaneous symmetry breaking of $B-L$ gauge symmetry. 
\item {\it Dark matter mass:\ }The mass square eigenvalue of dark matter field $\phi_{D }$ is given by,
\begin{equation}
m_{\phi_{DM}}^2=\mu_{D}^2+\frac{\lambda_{Dh}v^2}{2}+\frac{\lambda_{SD}v_{BL}^2}{2}. 
\end{equation}
In this work, we consider the  couplings $\lambda_{SD},\lambda_{Dh} $ to be very small $\sim 10^{-10}-10^{-13}$ to accommodate $\phi_{D }$ as  non-thermal dark matter. We also consider $\lambda_D$ of similar order $10^{-10}$, which suppresses any large contribution from $2 \to 4 $ processes, that could have brought the dark matter into kinetic and chemical equilibrium \cite{Arcadi:2019oxh}. Due to the choice of a small $\lambda_D$, its impact on the thermal correction of dark matter mass would be negligibly small. This also implies negligible impact of the phase transitions for our choices of dark matter masses which are in the range of a few GeV. To a good approximation, we therefore identify that the dark matter mass is primarily governed by the bare mass term, {\it i.e.,}  $m_{\phi_{DM}} \sim  \mu_{D}$ and ignore the thermal mass correction of the dark matter. 
\end{itemize}
Before finishing this section, we present a brief discussion  about the stability of the dark matter in this model. This is to note, that the dark matter does not acquire a vev in this model. However, since dark matter is charged under $B-L$ and the same symmetry is broken due to non-zero vev of $\mathcal{S}$ field, hence the dark matter will not be a stable dark matter for all values of $q_{DM}$.  As given in Table.~\ref{tabbml}, the dark matter candidate $\phi_{D }$ has charge $q_{DM}$ under $U(1)_{B-L}$. By choosing appropriate $q_{DM}$ with a value  $q_{DM}\neq \pm2n$ ($n \in \mathbb{Z}$ and $n\leq 4$), one  can avoid Yukawa interaction terms, such as, $\phi_{D }\bar{N}^c N$ and cubic and quartic interaction term, such as $\phi_{D }\mathcal{S}^2$ and  $\phi_{D }\mathcal{S}^3$ \cite{Rodejohann:2015lca}. Therefore, the decay of $\phi_{D }$ can be forbidden without invoking extra discrete symmetry in the model and hence $\phi_{D }$ can be the viable stable dark matter candidate. For dark matter in a different representation other than being $ SU(2)_L$ singlet, additional re-normalizable and non-renormalizable operators involving SM fields may present, which can further contribute to dark matter decay. This has been studied in \cite{Cirelli:2005uq}. 

In this work, we consider that $\phi_D$ is a dark matter with feeble interaction strengths (FIMP candidate). Therefore, in the early Universe, the state had negligible abundance and during reheating of the Universe  it was not in the thermal equilibrium. The dark matter $\phi_{D }$ has both $U(1)_{B-L}$ gauge and scalar interactions. The production of $\phi_{D }$ through gauge interactions are determined by gauge coupling $g_{BL}$ along with the charge  $q_{DM}$ of $\phi_{D}$ state,  the dark matter mass $m_{\phi_{DM }}$ and $ B-L$  gauge boson mass $m_{Z_{BL}}$. Here we primarily focus on the dark matter production via the scalar states and for this purpose the $q_{DM}$ is chosen to be sufficiently small, such that, the production of $\phi_{D }$  through gauge interactions becomes negligible. In the next section, we present a relative comparison between these two different production modes to justify our choice of parameters.

\section{Freeze-in Production of Dark Matter \label{freezeinprod}}	

\begin{figure}[h]
	\centering
	\includegraphics[width=10.0cm,scale=0.8]{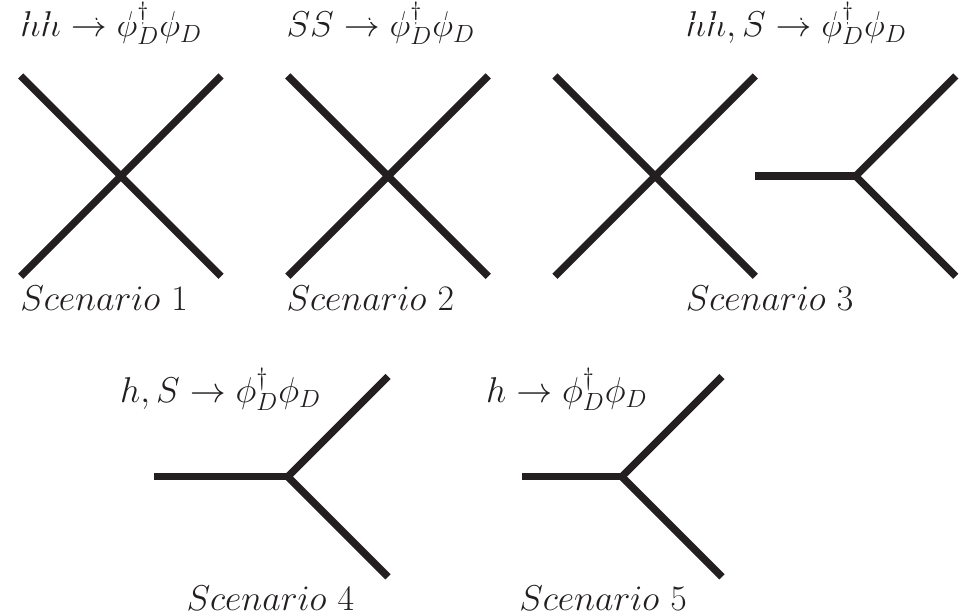}
	\caption{Schematic diagrams for different  dark matter production scenarios, {\it Scenario-(1-5)}. }\label{Fig1}
\end{figure}

\begin{figure}[h]
	\begin{center} 
		\includegraphics[width=0.65\linewidth,angle=-0]{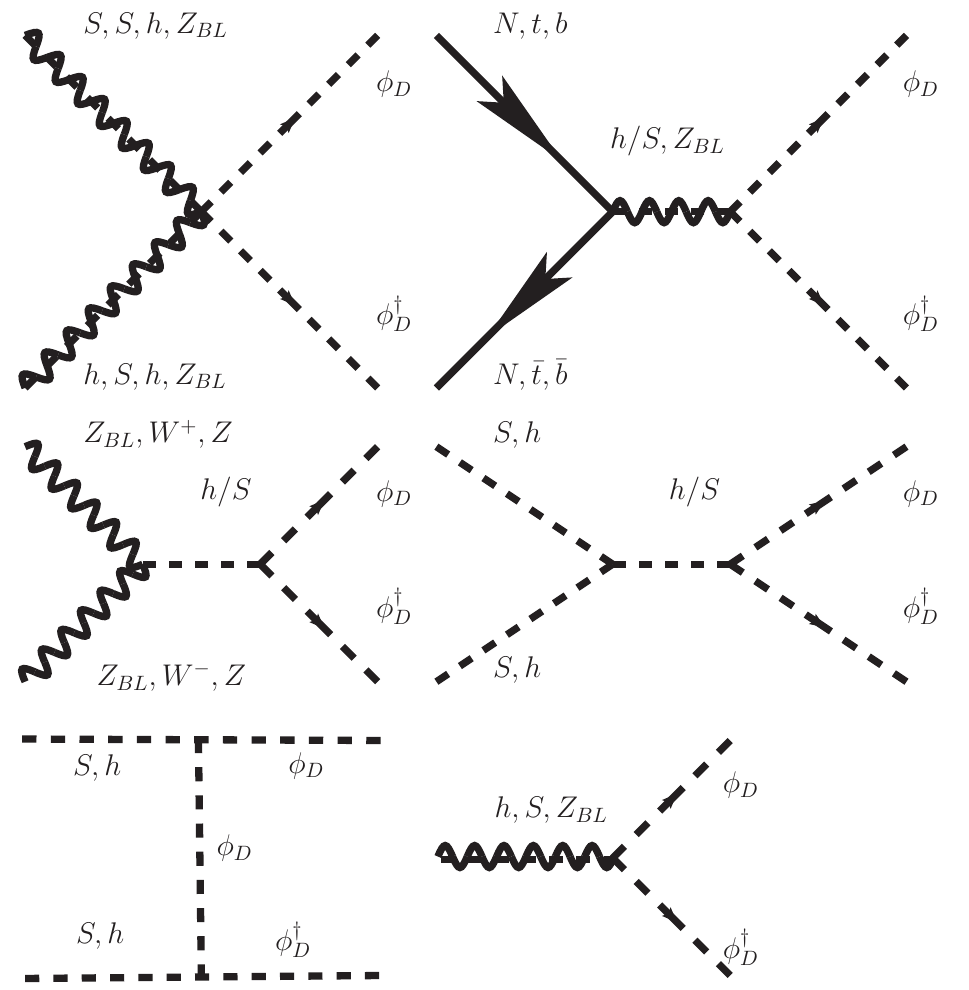}
		\caption{Production channels of dark matter $\phi_{D }$. The channels with $Z_{BL}$, as well as the $t$ channel contributions give negligible contributions for our scenario.  }\label{Feyndiag}
	\end{center}
\end{figure}

As outlined in the previous section, the dark matter particle $\phi_D$ has  feeble interactions with the SM particles, as well as,  other $B-L$ particles ($S,Z_{BL}$) present in this model.  Therefore, the state $\phi_D$ is not in thermal equilibrium, rather produced from the decays and annihilation  of SM and $B-L$ particles.  If kinematically allowed, the freeze-in production of dark matter is dominated by the decays of  SM and $B-L$ states which are in thermal equilibrium. The  production processes due to annihilation  give subdominant contributions to the relic density,  as often the contributions are suppressed by  additional couplings as well as propagators, along with the numerical factors arising from additional phase space integral. A non re-normalizable interactions between the dark matter and bath particles leads to UV freeze-in of dark matter which depends on the re-heating temperature of the Universe \cite{Hall:2009bx, Chen:2017kvz, Elahi:2014fsa, Biswas:2019iqm}. In this work, we do not have non-renormalizable interaction between the dark matter and bath particles. Rather, our scenario is similar to  IR freeze-in of dark matter,  where production of the dark matter dominates  at $T\approx M$ of the initial states  and  it is insensitive to the reheating temperature of the Universe. We consider both the decay and $2 \to 2$ annihilation contributions in the relic density.Feynman diagrams for the production processes of dark matter $\phi_{D }$ before and after EWSB are shown in Fig.~\ref{Feyndiag}.Depending on the primary production mechanism, we sub-divide the entire discussion in different {\it Scenarios}, and analyse the production of $\phi_D$ in detail. The schematic diagrams for these different scenarios have been shown in Fig.~\ref{Fig1}. 


\begin{itemize}
\item
\sn{1}: The dark matter is primarily produced from the annihilation of the SM Higgs boson. 
\item
\sn{2}: The dark matter is produced primarily from the annihilation  of the $B-L$ Higgs, with a sub-dominant contribution from the annihilation of the SM Higgs boson.
\item	
\sn{3}: The dark matter production is governed by the annihilation of the SM Higgs boson at an earlier epoch, but later dominated by the decay of the $B-L$ Higgs bosons.	
\item
\sn{4}: The dark matter production is governed by the decays of SM and  $B-L$ Higgs bosons.	
\item
\sn{5}: The dark matter is produced  mainly from the decay of SM  Higgs boson  with a sub-dominant contribution from the  $B-L$ Higgs boson. In the earlier epoch, the dark matter production is primarly governed by the  SM Higgs annihilation.	
\end{itemize}
\begin{figure}[b]
		\mbox{\subfigure[]{\includegraphics[width=0.52\linewidth,angle=-0]{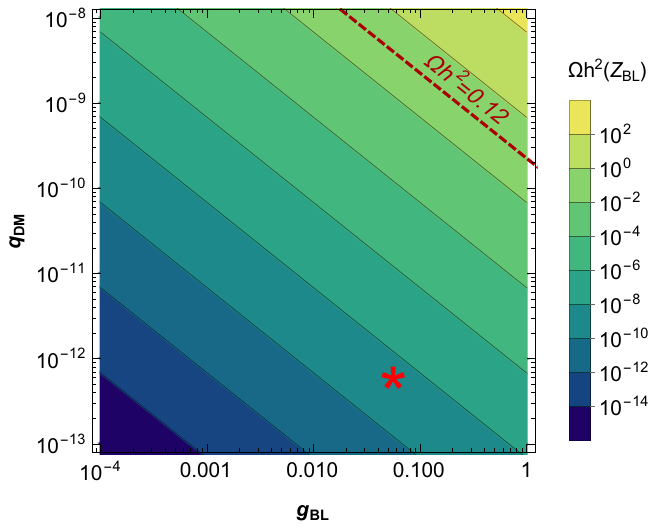}\label{k1}}
		\subfigure[]{\includegraphics[width=0.52\linewidth,angle=-0]{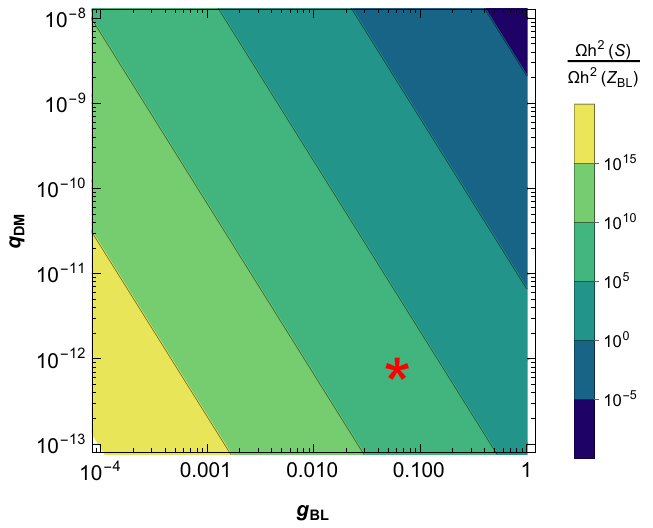}\label{k2}}}
		\caption{Fig.~\ref{k1} shows contours  of  relic abundance of $\phi_D$ in the $g_{BL}$ and  $q_{DM}$ plane, where production of dark matter is governed by gauge interaction. Fig.~\ref{k2} shows  contours of the  ratio of relic abundance of $\phi_{D }$ from $B-L$ Higgs boson $S$ and $Z_{BL}$ decay. The parameters chosen are as follows, $m_{\phi_{DM}}$=1\, GeV, $m_{Z_{BL}}$=5.5\, TeV, $m_{S}$=200\, GeV \ and $\lambda_{SD}=10^{-13}$. The red star in Fig.~\ref{k1} and Fig.~\ref{k2}  represents our benchmark point.}\label{Fig2}
\end{figure}
We explore each of these different scenarios in detail taking into account all the relevant contributions in the Boltzmann equation.  However, before focusing on the main study of this paper,  we bring the attention of the readers on a  comparative study between the $B-L$ gauge boson ($Z_{BL}$) contribution and $B-L$ scalar ($\mathcal{S}$) contribution to the dark matter relic density.  It is well known, that when the dark matter is gauged it would quickly thermalise due to potentially larger effective gauge coupling and charge associated with it. In our case such a phenomena can happen as dark matter can be copiously produced via $Z_{BL}$ decay and/or by the annihilation mediated by $Z_{BL}$ or via contact interaction. Such process can lead to overproduction of dark matter in the very early Universe,  and the only viable option to maintain the correct dark matter relic is freeze-out \cite{Bandyopadhyay:2017bgh, Rodejohann:2015lca}.  However, our goal for this article is to investigate the possibility of relativistic freeze-in scenario which compels us to choose a very small value of $q_{DM}$.\newline 
The number density of $\phi_{D }$ from $Z_{BL}$ decay  {\it i.e.,} $Z_{BL} \to \phi^*_{D} \phi_{D}$ and annihilation  processes {\it i.e.,} $f \bar{f} \to \phi^*_D \phi_{D}$  are calculated by the following Boltzmann equation,
	\begin{equation}
	\frac{d n_{\phi_{D }}}{dt}+ 3 H n_{\phi_{D }}= \Gamma_{Z_{BL} \rightarrow \phi^*_{D } \phi_{D }}+ \sum_{f=N,t,b}\Gamma_{\bar{f}f \rightarrow \phi^*_{D } \phi_{D }},
	\label{boltz1}
	\end{equation}
	where $\Gamma_{aa \rightarrow bb }$ and $\Gamma_{a \rightarrow bb }$  represent the reaction rates for annihilation and decays. 
	In comoving volume the above Boltzmann equation can be written as, 
	\begin{equation}\begin{aligned}
	\frac{dY_{\phi_{D }}}{d z}= \frac{z^{4}}{s(m_{\phi_{DM}}) H(m_{\phi_{DM }})} [\Gamma_{Z_{BL} \rightarrow \phi^*_{D } \phi_{D }}+ \sum_{f=N,t,b}\Gamma_{\bar{f}f \rightarrow \phi^*_{D } \phi_{D }}]. 
	\end{aligned}
	\end{equation}
	The relic abundance of $\phi_{D }$ here is mostly dominated by  $Z_{BL}$ decay and  is given by, 
	\begin{equation}
	\Omega h^2(Z_{BL})=\frac{m_{\phi_{DM}} s_0 Y_{\phi_{D }(\infty)}}{\rho_c / h^2}. 
	\end{equation}
	
Fig.~\ref{k1} represents the relic density contours where, we vary the charge $q_{DM}$ and the coupling $g_{BL}$.   This is to note, that the production of $\phi_{D }$ via $Z_{BL}$ mediated annihilation processes, {\it i.e.,} $\bar{f}f,\bar{N}^c N \rightarrow Z_{BL}\rightarrow \phi^*_{D }\phi_{D }$ are also kinematically allowed but  such processes are suppressed due to fourth power of $g_{BL},$ as well as large $Z_{BL}$ mass compared to $Z_{BL}\rightarrow \phi^*_{D }\phi_{D}$ process. Therefore, in Fig.~\ref{k1}, we ignore the contributions from the annihilation processes mediated by $Z_{BL}$. In the same plot, we also show the contour that satisfies the present relic density $\Omega h^2=0.12$ \cite{Ade:2015xua} by the red dashed line.  We can easily infer that for $q_{DM} \gsim 10^{-10}$, $Z_{BL} \to \phi^*_D \phi_D$ contribution alone attains the desired relic density. On the other hand, the star mark in  Fig.~\ref{k1} represents our chosen benchmark point, for which the $Z_{BL} \to \phi^*_D \phi_D$ decay gives negligible contribution in the relic density. For a fixed $g_{BL}$, as we increase $q_{DM}$, contribution  from $Z_{BL}$ decay in relic density will increase. For a very large $q_{DM} \sim 10^{-1}$, the DM will thermalise with the SM particles, and freeze-out scenario will be the viable option \cite{Rodejohann:2015lca}. \newline
A comparative study of  $S \to \phi^*_D \phi_D$  and  $Z_{BL} \to \phi^*_D \phi_D$ processes is presented in Fig.-\ref{k2} where, we show the ratio of the relic densities. For our choice of masses, as given in the caption of Fig.~\ref{Fig2}, the decay of both $B-L$ scalar $S$ and $Z_{BL}$ into two $\phi_D$ state  are kinematically  allowed. The ratio increases significantly with the decrease in $g_{BL}$ and $q_{DM}$, as can be explained from the following equation, 
\begin{equation}
\frac{\Gamma_{ S  \rightarrow \phi^*_{D } \phi_{D }}}{\Gamma_{ Z_{BL}  \rightarrow \phi^*_{D } \phi_{D }}}\propto \frac{\lambda_{SD}^2 m_{Z_{BL}}}{4 g_{BL}^4 q_{DM}^2 m_S}.\label{eq:ratiointeraction}
\end{equation}  
We  choose  $q_{DM}\approx 10^{-12}$ represented by the red star in Fig~\ref{k1} where it is evident that the production of $\phi_{D }$ through gauge interaction is negligible and thus we neglect contribution from the $B-L$ gauge interaction in the rest of the paper. Even if we consider $Z_{B-L}$ mass  different from 5.5 TeV \cite{CMS:2019tbu}, as long as we are choosing a sufficiently  small $q_{DM}$,  production of dark matter from the scalar sector will continue to dominate. We focus on the production of the dark matter from  decay and annihilation of the scalars in  relativistic freeze-in scenarios. The effect of SM fields (fermions, gauge bosons) are also taken into account via the interactions which are operative after EWSB. As discussed in the previous section, for our analysis we consider that the re-heating temperature of the Universe is same as the temperature at which $B-L$ symmetry breaks down. To evaluate dark matter number density,  we therefore perform the analysis in the $B-L$ broken phase. 
\newline
Below, we present a detailed discussion of the different scenarios, where we numerically solve the Boltzmann equation and evaluate the relic density.  In doing so, we consider different decay $a \to \phi^*_D \phi_D$ and annihilation/co-annihilation processes  $aa, ab \to \phi^*_D \phi_D$. 
	 	 
\subsection{\it{Scenario-1}:}
				\begin{table}[h]
					\centering
					\renewcommand\arraystretch{1.1}
					\begin{tabular}{c| c  c  c | c  c c c c}
						\hline\hline
						\multirow{2}{*}{ \textit{Scenario}  } 	& \multicolumn{3}{c}{Masses in GeV} & \multicolumn{5}{|c}{Couplings} \\
						& $m_S$ & $m_N$ & $m_{\phi_{DM}}$  & $y_N$ & $\lambda_{SD}$ & $\lambda_{Sh}$ & $\lambda_{NS}$& $\lambda_{Dh}$\\ \hline
						\textit{1} & 200  & 300 & 250 & $10^{-7}$ & $5.0\times10^{-12}$ & $6\times 10^{-6}$&$0.053$ &$1.6 \times10^{-11}$\\
						\hline\hline
					\end{tabular}
					\caption{The choices of masses and couplings for  {\it  Scenario-1}.  }\label{Table1}
				\end{table}
In this scenario, the dark matter production primarily occurs via SM Higgs boson annihilation {\it i.e.,} $h h \rightarrow \phi^*_{D } \phi_{D }$. We adopt a relativistic freeze-in framework for the evaluation of the relic density. The contribution of $S S \rightarrow \phi^*_{D } \phi_{D }$ is although allowed but small in attaining the correct dark matter relic. Since $m_{\phi_{DM}} > m_h/2 , m_S/2$ the  decay contributions from  the SM and $B-L$ Higgs  bosons are absent. The choices of masses and coupling used in the numerical analysis, are shown in Table.~\ref{Table1}. The Boltzmann equation for the production of $\phi_{D }$ in this scenario is given by Eq.~\eqref{BMsc1},
\bea\label{BMsc1}
&&\frac{d n_{\phi_{D }}}{dt}+ 3 H n_{\phi_{D }}=(4-3\theta (T_{EW}-T)) \Gamma_{h h \rightarrow \phi_{D }^\dagger \phi_{D }}+\Gamma_{S S  \rightarrow \phi_{D }^\dagger \phi_{D }}+\Gamma_{N N  \rightarrow \phi_{D }^\dagger \phi_{D }}\\
& &+\theta (T_{EW}-T) \left[\Gamma_{h S  \rightarrow \phi_{D }^\dagger \phi_{D }}+\Gamma_{W^+ W^- \rightarrow \phi_{D }^\dagger \phi_{D }}+\Gamma_{Z Z \rightarrow \phi_{D }^\dagger \phi_{D }}+\Gamma_{b \bar{b}  \rightarrow \phi_{D }^\dagger \phi_{D }}+\Gamma_{t \bar{t}  \rightarrow \phi_{D }^\dagger \phi_{D }}\right],\nn
\eea
where $\Gamma_{aa \rightarrow bb }$  and   $\Gamma_{a \rightarrow bb }$  are the rates of the annihilation and decay processes for the respective channels. In a comoving volume the above Boltzmann equation can be read in terms of the yield as described in Eq.~\eqref{yeildsc1},
\bea\label{yeildsc1}
&&\frac{dY_{\phi_{D }}}{d z}=\ \frac{z^{4}}{s H} \Big[(4-3\theta (z-z_{EW})) \Gamma_{h h \rightarrow \phi_{D }^\dagger \phi_{D }}+\Gamma_{S S  \rightarrow \phi_{D }^\dagger \phi_{D }}+\Gamma_{N N  \rightarrow \phi_{D }^\dagger \phi_{D }}\\
&& + \theta (z-z_{EW}) \left[\Gamma_{h S  \rightarrow \phi_{D }^\dagger \phi_{D }}+\Gamma_{W^+ W^- \rightarrow \phi_{D }^\dagger \phi_{D }}+\Gamma_{Z Z \rightarrow \phi_{D }^\dagger \phi_{D }}+\Gamma_{b \bar{b}  \rightarrow \phi_{D }^\dagger \phi_{D }}+\Gamma_{t \bar{t}  \rightarrow \phi_{D }^\dagger \phi_{D }}\right]\Big].\nn
\eea
where the number density $n$ is related with the entropy density $s$ as $n={Y}/{s}$ and $H$ is the Hubble's constant. Before EWSB, all four degrees of freedom of the SM Higgs doublet contribute to $\Phi \Phi \to \phi^*_D \phi_D$ leading to four times enhancement in the relic density as compared to $hh \to \phi^*_D \phi_D$ after EWSB (green line). The expressions of the different reaction rates are given in  Appendix.~\ref{appen2}, where we have closely followed the approach of \cite{Lebedev:2019ton, Arcadi:2019oxh}. The annihilation processes $h h, S S, N N$ are always open while the other SM annihilation processes unlatch only after EWSB.\newline
\begin{figure}[t]
	\begin{center}
		\mbox{\subfigure[]{\includegraphics[width=0.5\linewidth,angle=-0]{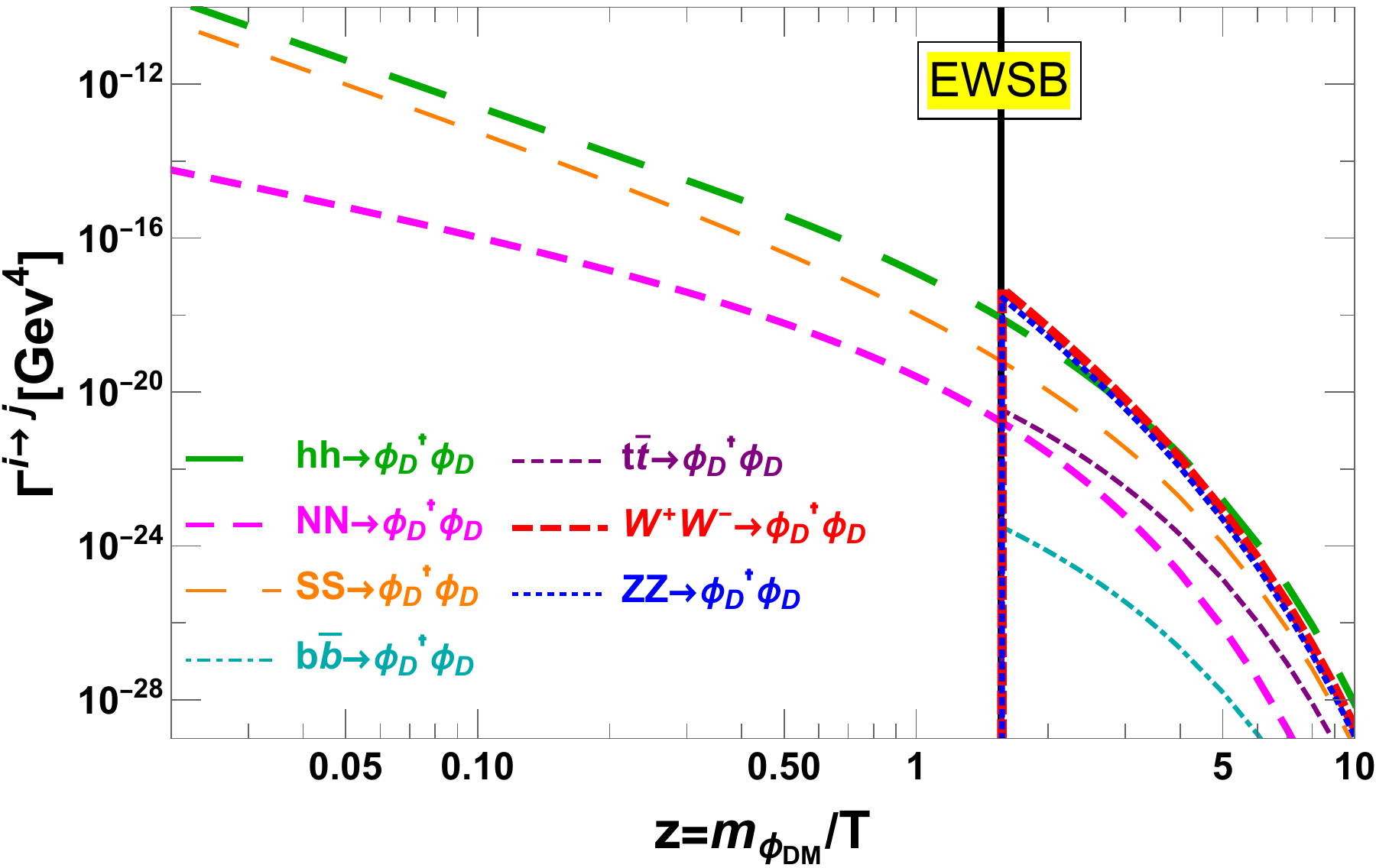}\label{f1}}
			\subfigure[]{\includegraphics[width=0.5\linewidth,angle=-0]{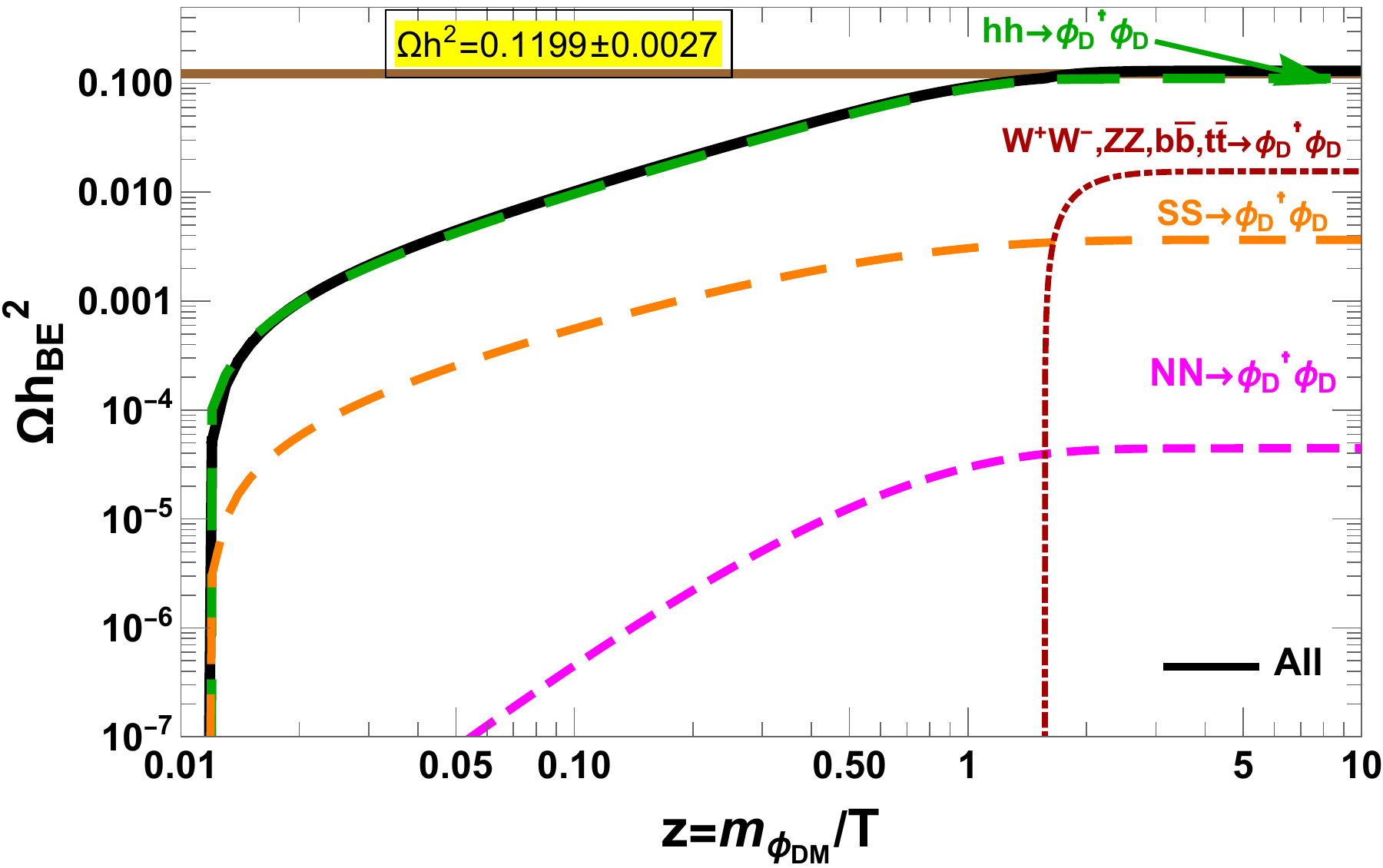}\label{f2}}}
		\mbox{\subfigure[]{\includegraphics[width=0.5\linewidth,angle=-0]{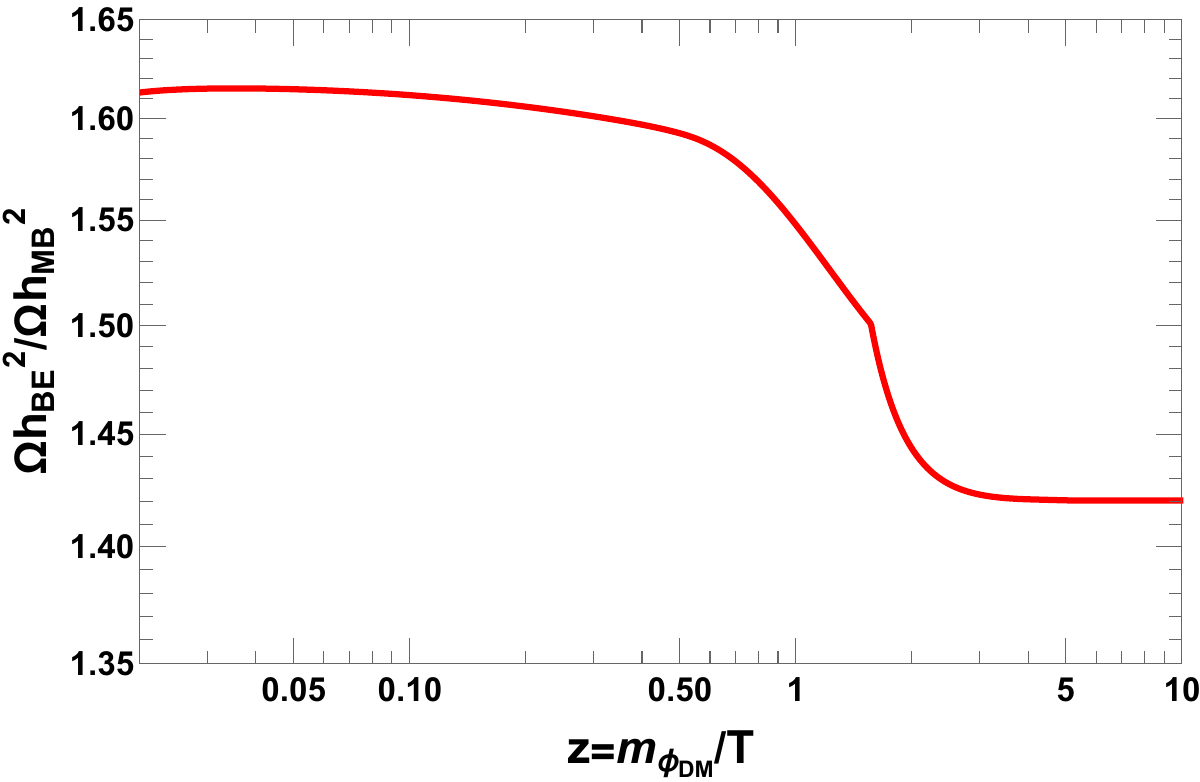}\label{f3}}}
		\caption{The figures correspond to {\it Scenario-1}. Fig.~\ref{f1} shows the relativistic reaction rates for the process $ h h \to \phi_D^\dagger \phi_D$ and other relevant processes. Fig.~\ref{f2} shows the individual contributions to the relic density, and the total  relic density. The brown horizontal line represents the present experimentally measured relic density \cite{Ade:2015xua}. Fig.~\ref{f3} shows the relative enhancement in the relic density with respect to Maxwell-Boltzmann distribution.}\label{Fig3}
	\end{center}
\end{figure}
In Fig.~\ref{f1} we show the relativistic rates for different annihilation processes $h h, \,S S, \, N N  \to \phi^*_D \phi_D$ as well as $W^{+} W^{-} , \,Z Z,\, t \bar{t} \to \phi^*_D \phi_D$ including the thermal correction of the SM Higgs boson mass. It is evident from Fig.~\ref{f1}  that $ h h \to \phi^*_D \phi_D$ is the most dominant mode for almost all values of $z$ while the other SM annihilation processes contribute only after EWSB. The $B-L$ Higgs boson contribution $SS \to \phi^*_D \phi_D$ (orange line) is large but subdominant. In Fig.~\ref{f2} we show the evolution of the dark matter relic density which attains freeze-in at the temperature of 160 GeV, and is dominated mainly by  $h h \to \phi^*_D \phi_D$ annihilation. The other observations are listed as follows:

\begin{itemize}
\item The Higgs annihilation processes $ \Phi \Phi, \, h h \to \phi^*_D \phi_D$,\footnote{$\Phi$ is the SM Higgs doublet and $h$ represents the SM Higgs field after EWSB. In Fig.~\ref{f1}, we do not maintain this distinction, rather represent the SM Higgs doublet(before EWSB) and SM Higgs field(after EWSB) by $h$ only. We show the contribution from one massive degree of freedom in $hh \to \phi_{D }^*\phi_{D }$. In Fig.~\ref{f2}, all the four contributions (before EWSB) and one contribution (after EWSB) have been considered.} is mostly dominated by the contact four point diagram, given in Fig.~\ref{Fig1}. However, for our numerical analysis contributions from all relevant diagrams (mediated via $h,S$) are taken into account. The cross-sections for the processes are listed in appendix.~\ref{appen1}. It is worth mentioning that due to the choice of the dark matter mass none of the annihilation process contains any resonant production. The $t$-channel diagram gives negligible contribution, and hence, is not considered. 
\item
Similar to the previous case, for the annihilation channel $ S S \to \phi^*_D \phi_D$, dominant contribution arises from the contact term. 
\item The annihilation rate of $NN \to \phi^*_D \phi_D$ is much suppressed as compared to $\Phi \Phi, \, hh \to \phi^*_D \phi_D$ due to additional couplings with $S$ and the corresponding propagators.
\item
The SM particles annihilate into dark matter state $t \bar{t},\, W^+ W^-, \,ZZ, b \bar{b} \to \phi^*_D \phi_D$. These processes are mediated primarily by the SM Higgs, and hence only open up after EWSB. Due to the small mixing between SM and $B-L$ Higgs bosons the contributions from $B-L$ Higgs boson in these processes are very small. As already mentioned, the choices of dark matter mass restrains to have any resonant annihilation via the SM Higgs mediation.
\end{itemize}
Fig.~\ref{f3} depicts the relative enhancement in the relic density obtained using BE distribution as compared to MB distribution, which is as significant as $\sim 1.42-1.62$. The relative interaction strength $\frac{\Gamma_{BE}}{\Gamma_{MB}}$ also varies accordingly with $z$.
		
\subsection{\it Scenario-2}
Unlike the previous scenario,  the $\phi_D$ production is governed primarily by annihilation  of the $B-L$ Higgs ($S S \rightarrow \phi^*_{D } \phi_{D }$) with  sub-dominant contributions from  $ \Phi \Phi,\, h h  \rightarrow \phi^*_{D } \phi_{D }$. The larger production from  $B-L$ scalar annhilation occurs due to a larger $\lambda_{SD}$ compared to $\lambda_{Dh}$, as can be seen from Table~\ref{Table2}. Similar to the \sn1, here also $h$ and $S$ decays are kinematically forbidden. The Boltzmann equation in this case would be the same as  Eq.~\eqref{BMsc1} and so is the yield equation  {\it i.e.,} Eq.~\eqref{yeildsc1}.
\begin{table}[t]
	\centering
	\renewcommand\arraystretch{1.1}
	\begin{tabular}{c| c  c  c | c  c c c c}
		\hline\hline
		\multirow{2}{*}{ \textit{Scenario}  } 	& \multicolumn{3}{c}{Masses in GeV} & \multicolumn{5}{|c}{Couplings} \\
		& $m_S$ & $m_N$ & $m_{\phi_{DM}}$  & $y_N$ & $\lambda_{SD}$ & $\lambda_{Sh}$ & $\lambda_{NS}$& $\lambda_{Dh}$\\ \hline
		\textit{2} & 200  & 300 & 150 & $10^{-7}$ & $3.0 \times10^{-11}$ & $6 \times 10^{-6}$&$0.053$ &$7.5 \times10^{-12}$\\
		\hline\hline
	\end{tabular}		
	\caption{The choices of masses and couplings for  {\it Scenario-2}.  }\label{Table2}	
\end{table}
\begin{figure}[t]
	\begin{center}
		\mbox{\subfigure[]{\includegraphics[width=0.49\linewidth,angle=-0]{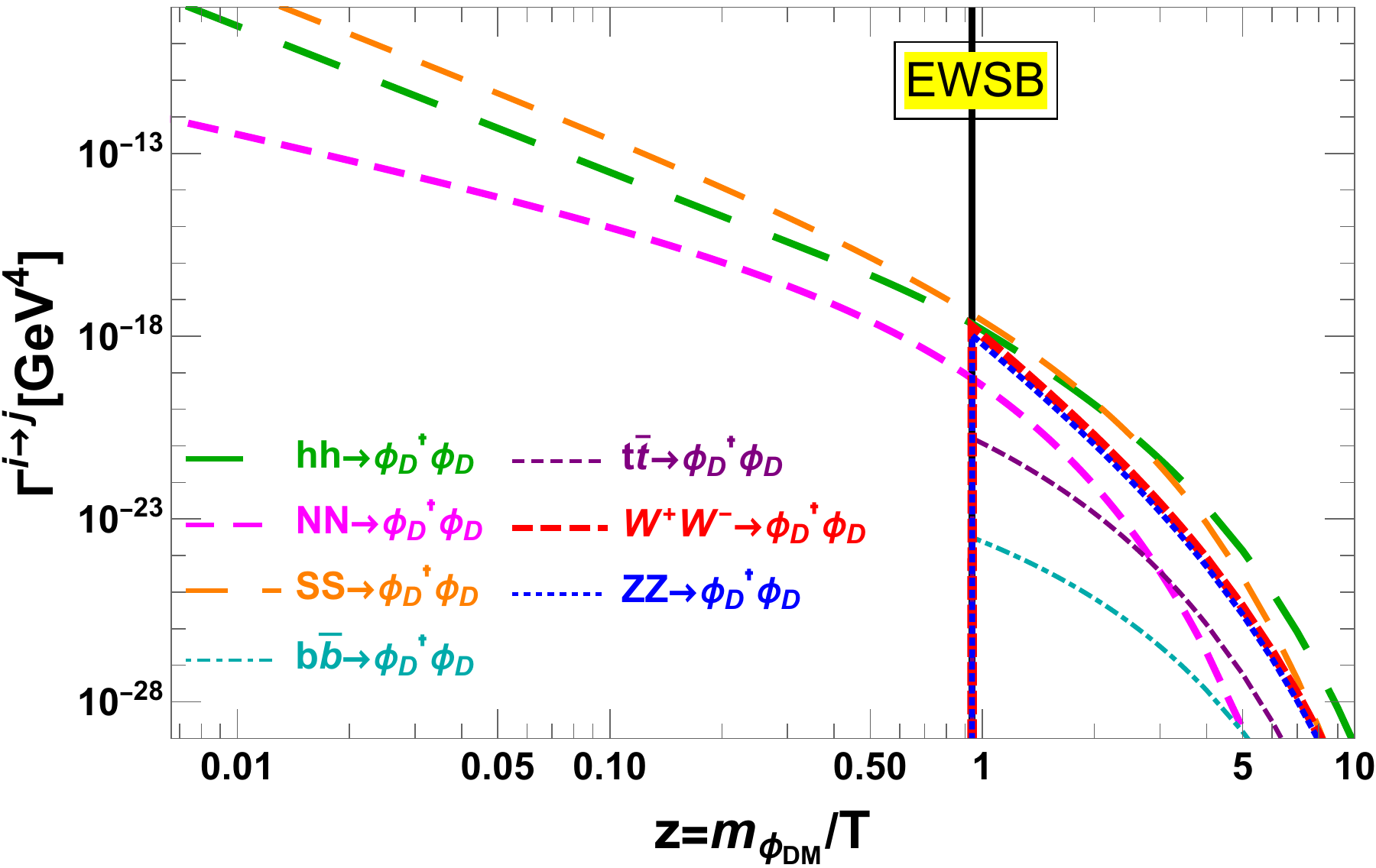}\label{ff1}}
			\subfigure[]{\includegraphics[width=0.49\linewidth,angle=-0]{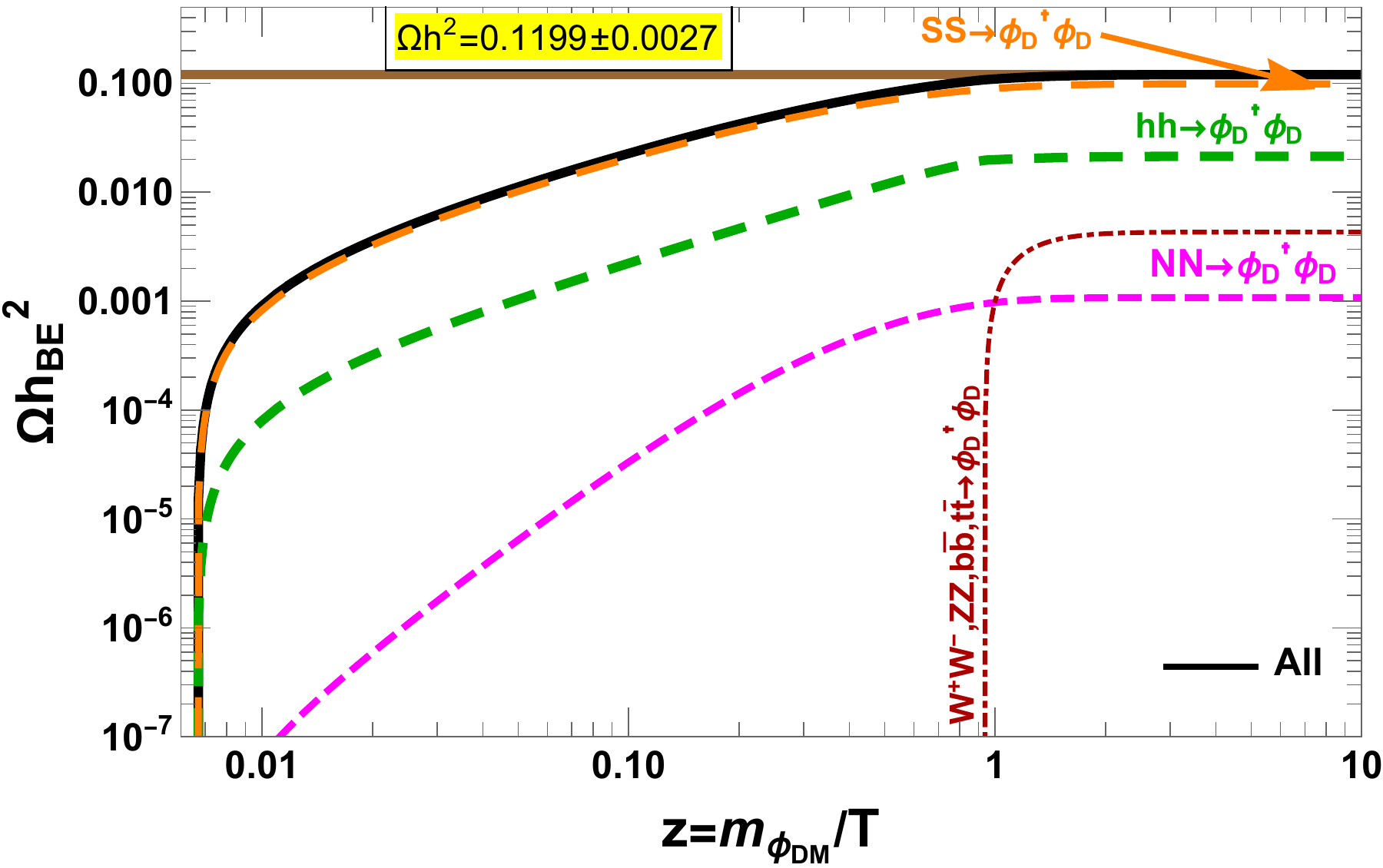}\label{ff2}}}
		\mbox{\subfigure[]{\includegraphics[width=0.49\linewidth,angle=-0]{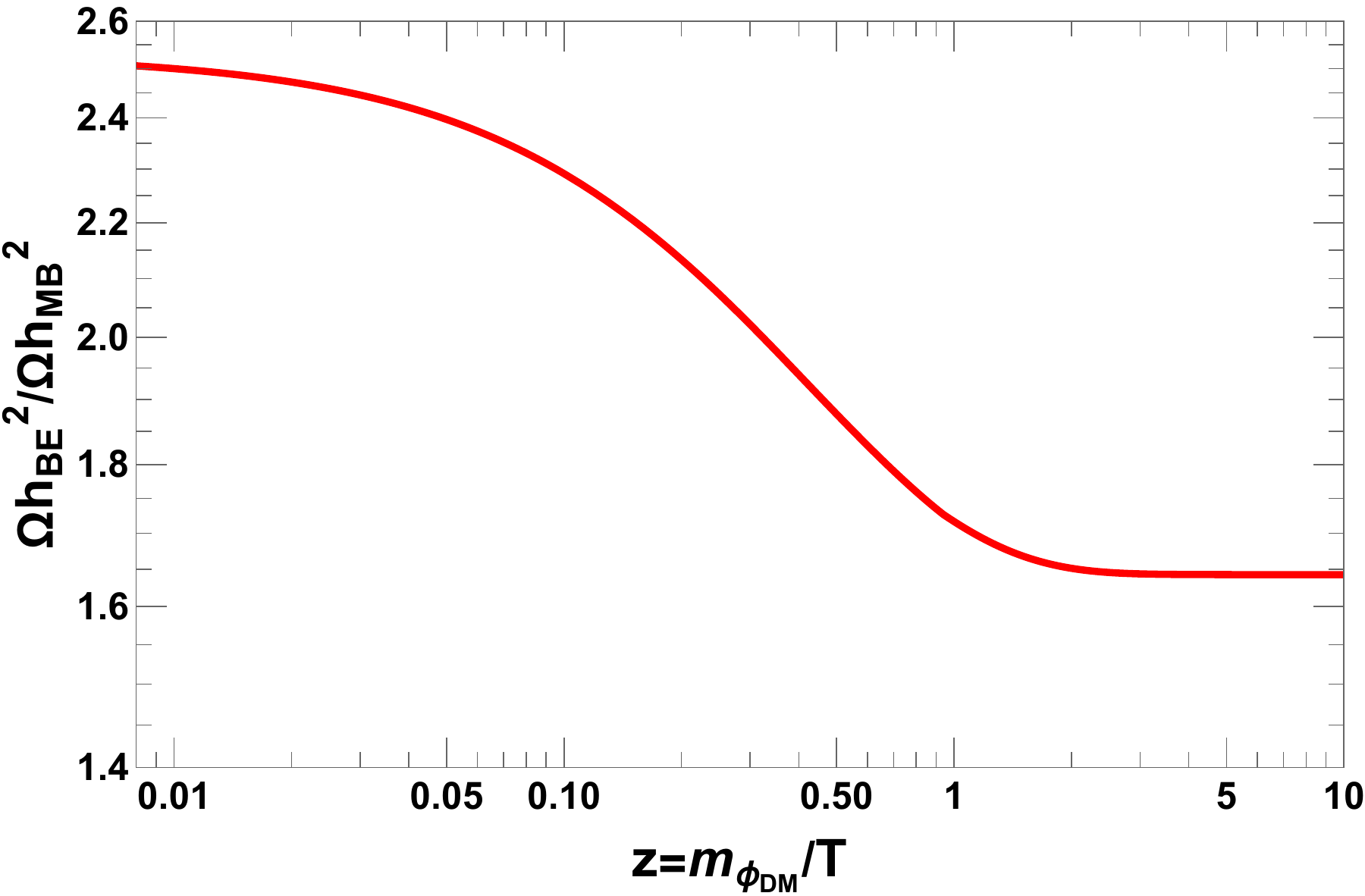}\label{ff3}}}
		\caption{The figures correspond to {\it Scenario-2}. Fig.~\ref{ff1} shows the relativistic reaction rates for different process corresponding to production of $\phi_{D }$. Fig.~\ref{ff2} shows the individual contributions to the relic density, and the total  relic density. The brown horizontal line represents the present experimentally measured relic density \cite{Ade:2015xua}. Fig.~\ref{ff3} shows the relative enhancement in the relic density  with respect to Maxwell-Boltzmann distribution.}\label{Fig4}
	\end{center}
\end{figure}
Similar to the previous case here also all the contributions coming from $h h, \, SS, NN\to \phi^*_D \phi_D$ are taken into account for the numerical analysis. The results are manifested in Fig.~\ref{Fig4}. In Fig.~\ref{ff1}, we show the relativistic rates for different processes corresponding to production of $\phi_{D }$. It can be seen that $S S \to \phi^*_D \phi_D$(orange line)  is the most-dominant, where this channel is governed by the contact interaction. $h h \rightarrow \phi^*_{D } \phi_{D }$ rate (green line) is the second dominant and its contribution is only 16\% to the dark matter relic. Due to the choice of the mass of dark matter, neither of the above two channels entail any resonance enhancement.
 
In Fig.~\ref{ff2}, we show the production of $\phi_{D }$ from different processes. In the present scenario, the dark matter freeze-in occurs at the temperature of 150 GeV. We can see that the $N N  \rightarrow \phi^*_{D } \phi_{D }$ annihilation mode is suppressed  due to additional couplings with $S, Z_{BL}$ and the corresponding propagators. The SM annihilation to dark matter, {\it i.e.,} $W^+ W^-, \,Z Z,\, t \bar{t}, \, b \bar{b} \to \phi^*_D \phi_D$ starts only after EWSB and are mediated via off-shell $h, S$. These processes contribute $\sim 3.3\%$ only.\newline
Fig.~\ref{ff3} depicts the relative enhancement of relic abundance using BE distribution over the MB distribution. We can see that the dark matter production is dominated by annihilation and the relative enhancement in the relic density using BE distribution is quite significant $\sim 2.5-1.65$.
\subsection{\it Scenario-3}\label{sc3}
\begin{figure}[t]
		\begin{center}
			\mbox{\subfigure[]{\includegraphics[width=0.56\linewidth,angle=-0]{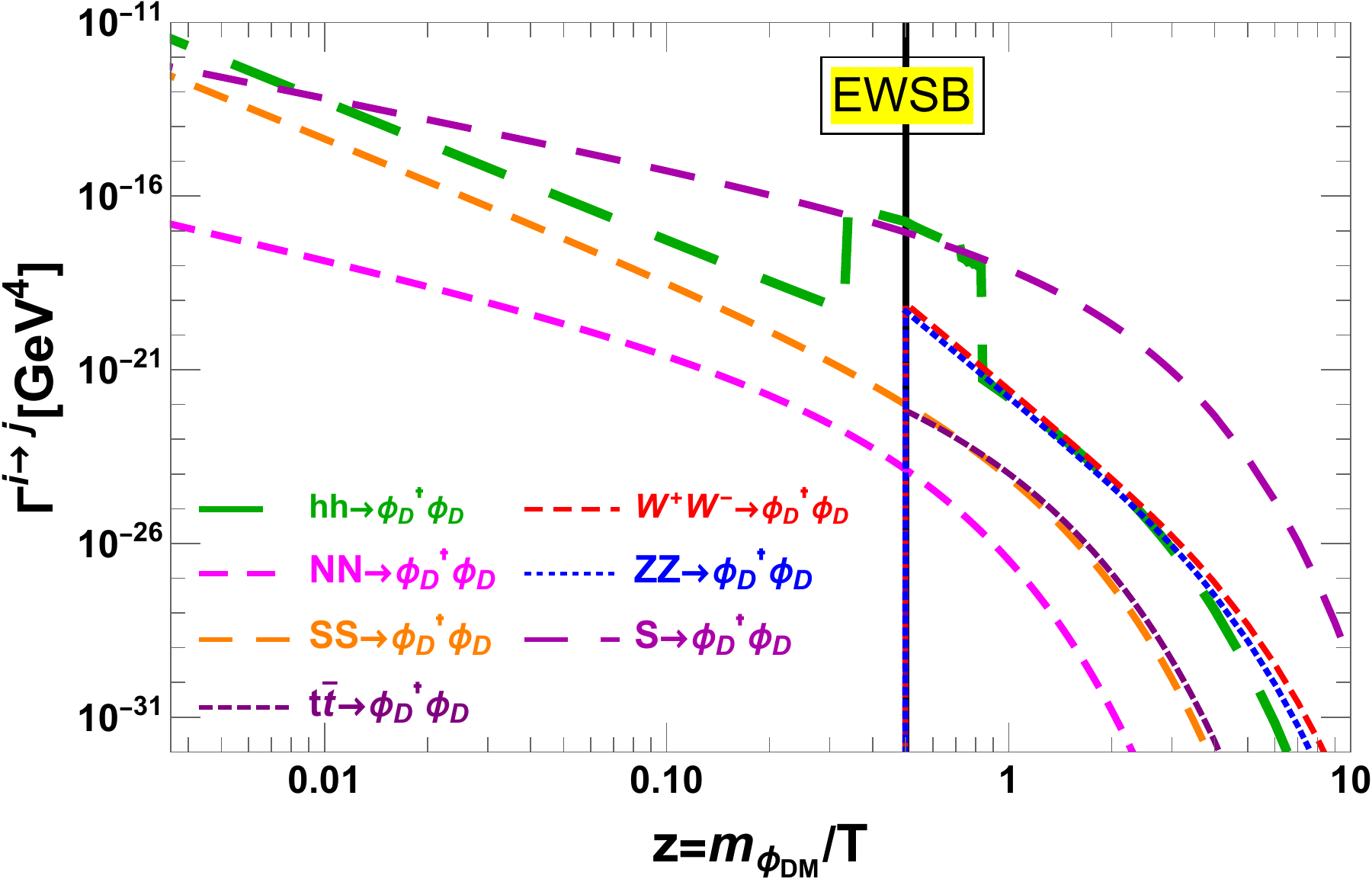}\label{a1}}
				\subfigure[]{\includegraphics[width=0.56\linewidth,angle=-0]{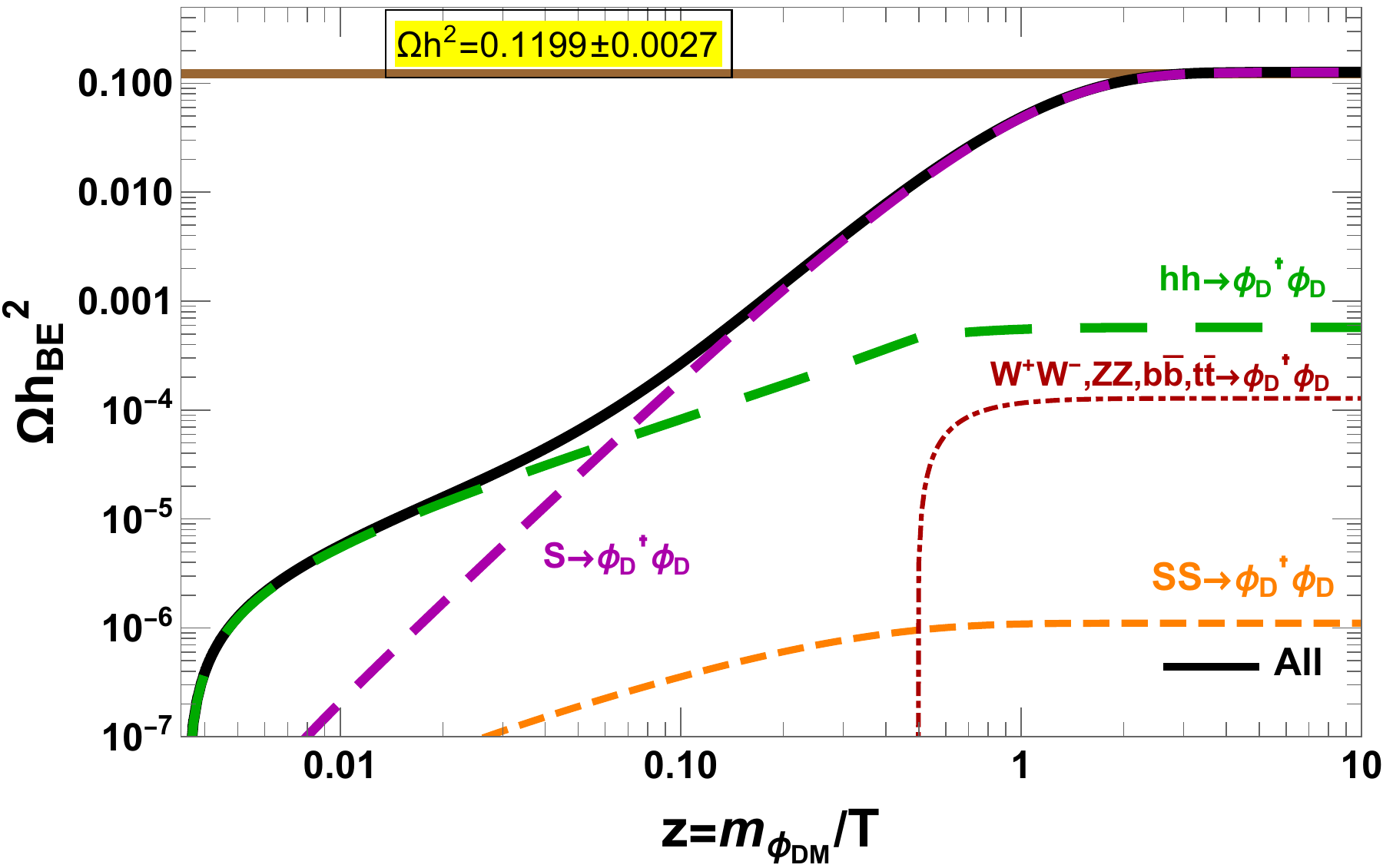}\label{a2}}}
			\mbox{\subfigure[]{\includegraphics[width=0.56\linewidth,angle=-0]{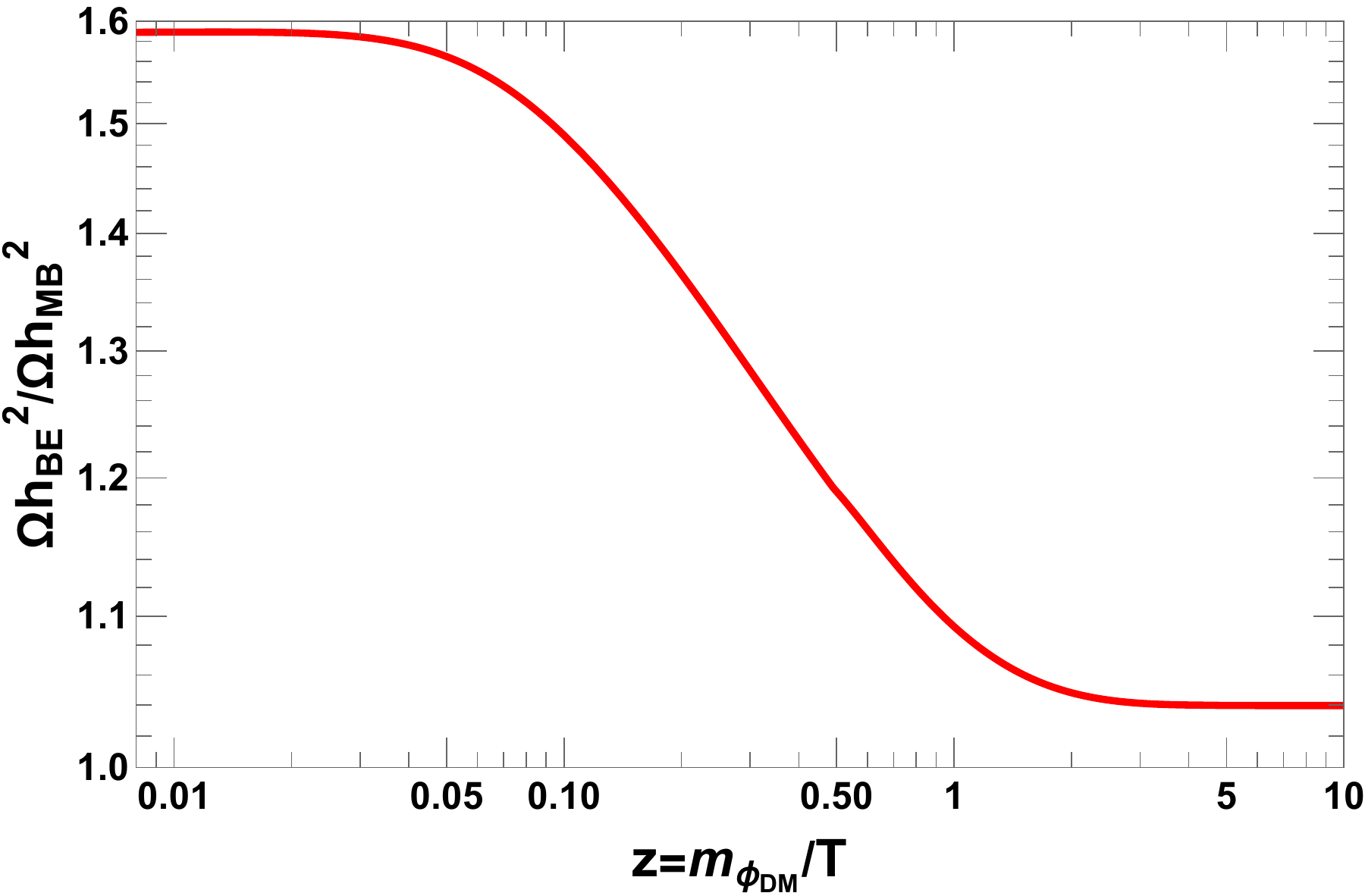}\label{a3}}}
			\caption{The figures correspond to {\it Scenario-3}. Fig.~\ref{a1} shows the relativistic reaction rates for different process corresponding to production of $ \phi_D$, and $S  \to \phi_D \phi_D$. Fig.~\ref{a2} shows the individual contributions to the relic density, and the total  relic density. The brown horizontal line represents the present experimentally measured relic density \cite{Ade:2015xua}. Fig.~\ref{a3} shows the relative enhancement in the relic density  with respect to Maxwell-Boltzmann distribution.}\label{Fig5}
		\end{center}
	\end{figure}

	\begin{table}[h]
	\centering
	\renewcommand\arraystretch{1.1}
	\begin{tabular}{c| c  c  c | c  c c c c}
		\hline\hline
		\multirow{3}{*}{ \textit{Scenario}  } 	& \multicolumn{3}{c}{Masses in GeV} & \multicolumn{5}{|c}{Couplings} \\
		& $m_S$ & $m_N$ & $m_{\phi_{DM}}$  & $y_N$ & $\lambda_{SD}$ & $\lambda_{Sh}$ & $\lambda_{NS}$& $\lambda_{Dh}$\\ \hline
		\textit{3} & 200  & 300 & 80 & $10^{-7}$ & $1.28 \times10^{-13}$ & $6 \times 10^{-6}$&$0.053$ &$1.414\times10^{-12}$\\
		\hline\hline
	\end{tabular}	
	\caption{The choices of masses and couplings for  {\it Scenario-3}.  }\label{Table3}		
\end{table}
Along with the SM and $B-L$ Higgs boson annihilation modes, in this case decay of $B-L$ Higgs boson $S$ becomes kinematically open due to the choice of parameter, shown in Table~\ref{Table3}. The decay $h \to \phi^*_D \phi_D$ however still remains forbidden. The Boltzmann equation contains the decay contribution as well, and can be written as, 
\bea\label{BMsc3}
&&\frac{d n_{\phi_{D }}}{dt}+ 3 H n_{\phi_{D }}=(4-3\theta (T_{EW}-T)) \Gamma_{h h \rightarrow \phi_{D }^\dagger \phi_{D }}+\Gamma_{S S  \rightarrow \phi_{D }^\dagger \phi_{D }}+\Gamma_{N N  \rightarrow \phi_{D }^\dagger \phi_{D }}+\Gamma_{S  \rightarrow \phi_{D }^\dagger \phi_{D }} \nn\\
&&+ \theta (T_{EW}-T) \left[\Gamma_{h S  \rightarrow \phi_{D }^\dagger \phi_{D }}+\Gamma_{W^+ W^- \rightarrow \phi_{D }^\dagger \phi_{D }}+\Gamma_{Z Z \rightarrow \phi_{D }^\dagger \phi_{D }}+\Gamma_{b \bar{b}  \rightarrow \phi_{D }^\dagger \phi_{D }}+\Gamma_{t \bar{t}  \rightarrow \phi_{D }^\dagger \phi_{D }}\right]. 
\eea
The yield equation in the comoving volume is given by,
\bea\label{yieldsc3}
\frac{dY_{\phi_{D }}}{d z}&=& \frac{z^{4}}{sH} \Big[(4-3\theta (z-z_{EW})) \Gamma_{h h \rightarrow \phi_{D }^\dagger \phi_{D }}+\Gamma_{S S  \rightarrow \phi_{D }^\dagger \phi_{D }}\nn\\
&&+\Gamma_{N N  \rightarrow \phi_{D }^\dagger \phi_{D }}+\Gamma_{S  \rightarrow \phi_{D }^\dagger \phi_{D }} + \theta (z-z_{EW}) \Big[\Gamma_{h S  \rightarrow \phi_{D }^\dagger \phi_{D }}\\
&&+\Gamma_{W^+ W^- \rightarrow \phi_{D }^\dagger \phi_{D }}+\Gamma_{Z Z \rightarrow \phi_{D }^\dagger \phi_{D }}+\Gamma_{b \bar{b}  \rightarrow \phi_{D }^\dagger \phi_{D }}+\Gamma_{t \bar{t}  \rightarrow \phi_{D }^\dagger \phi_{D }}\Big]\Big]. \nn
\eea
As before, here also we consider all possible annihilation processes $h h, \,S S, NN,\, hS  \to \phi^*_D \phi_D$ and decay $S \to \phi^*_D \phi_D$, along with other SM processes. The results are summerised  in Fig.~\ref{Fig5}. Fig.~\ref{a1} represents the relativistic rates for the different processes corresponding to the production of $\phi_{D }$. Fig.~\ref{a2} shows the evolution of $\phi_{D }$. The annihilation  mode $h h \to \phi^*_D \phi_D$ (green line) remains most efficient in the production of the dark matter for $z<0.02$. After this $S \to \phi^*_D \phi_D$ (purple line) takes over and remains the dominant mode until dark matter freezes-in at  the temperature of 40 GeV. Since the $S$ decay is open, an abrupt  increase in $\lambda_{ SD}$ can cause an overproduction of $\phi_{D }$. Around EWSB where the Higgs mass falls below  $100\,  \textrm{GeV} $, $s$-channel resonance occurs in the process $h h  \rightarrow \phi^*_{D } \phi_{D } $ (green bump) which enhances the production rate.\newline
Here we elaborate on an important point about our calculation which resolves the possible over counting. We take into account all possible contributions in the evaluation of  $h h  \rightarrow \phi^*_{D } \phi_{D }$ process, namely  the contact term, on-shell and off-shell contributions mediated by $h,\,S$. The $S$ mediated $s$-channel diagram encounters  a resonance  around  EWSB. In this case the production of $\phi_{D }$ must be computed by subtracting the  on-shell $S$ exchange contribution to avoid any over-counting, as in the Boltzmann equation this effect has already been taken into account by the $S$ decay contribution separately (Eq.~\eqref{BMsc3}) \cite{Belanger:2018ccd}.  The on-shell contribution  due to $S$ mediation  is given by,
	\begin{equation}
	\Gamma^{on-shell}_{h h \rightarrow \phi_{D }^\dagger \phi_{D }}=\Gamma_{h h \rightarrow S}\textrm{BR}(S \to \phi_{D}^\dagger \phi_{D }).
	\end{equation}
Therefore, only contact term and off-shell contribution $hh\rightarrow S^* \to  \phi_{D }^\dagger \phi_{D }$ (also  $hh\rightarrow h^* \to  \phi_{D }^\dagger \phi_{D }$) are taken into account for the $\phi_{D }$ production in this scenario.\newline
Fig.~\ref{a3} depicts the relative enhancement of relic abundance using BE and MB distribution. One can see that at a very early epoch, where  the dark matter production was dominated only by SM Higgs boson $h$ annihilation,  the ratio is very significant around $\sim 1.6$. At the later epoch, when the  production of $\phi_{D }$ is dominated by the $S$ decay, we find that the enhancement is about 1.04.
\subsection{\it Scenario-4}\label{sc4}
\begin{table}[h]
	\centering
	\renewcommand\arraystretch{1.1}
	\begin{tabular}{c| c  c  c | c  c c c c}
		\hline\hline
		\multirow{2}{*}{ \textit{Scenario}  } 	& \multicolumn{3}{c}{Masses in GeV} & \multicolumn{5}{|c}{Couplings} \\
		& $m_S$ & $m_N$ & $m_{\phi_{DM}}$  & $y_N$ & $\lambda_{SD}$ & $\lambda_{Sh}$ & $\lambda_{NS}$& $\lambda_{Dh}$\\ \hline
		\textit{4} & 200  & 300 & 1 & $10^{-7}$ & $6.65 \times10^{-13}$ & $ 6\times10^{-6}$&$0.053$ &$8.6\times10^{-12}$\\
		\hline\hline
	\end{tabular}			
	\caption{The choices of masses and couplings for  {\it Scenario-4}.  }\label{Table4}
\end{table}
This is the most generic scenario where along with different annihilation processes, the decays of both the Higgs bosons $h, S \to \phi^*_D \phi_D$ are kinematically allowed. The chosen benchmark points are tabulated in Table~\ref{Table4}. Unlike previous cases, the dark matter in this scenario is very light $m_{\phi_{DM}}=1$  GeV. As we will show in the subsequent discussion, the decay of Higgs bosons $h,S$ give the most dominant contribution in the relic density. The  most generic Boltzmann equation involved in this case  has the following form: 
\bea
\frac{d n_{\phi_{D }}}{dt}+ 3 H n_{\phi_{D }}&=&(4-3\theta (T_{EW}-T)) \Gamma_{h h \rightarrow \phi_{D }^\dagger \phi_{D }}+\Gamma_{S S  \rightarrow \phi_{D }^\dagger \phi_{D }}+\Gamma_{N N  \rightarrow \phi_{D }^\dagger \phi_{D }}+\Gamma_{S  \rightarrow \phi_{D }^\dagger \phi_{D }}\nn\\
&&+ \theta (T_{EW}-T) \Big[\Gamma_{h  \rightarrow \phi_{D }^\dagger \phi_{D }}+\Gamma_{h S  \rightarrow \phi_{D }^\dagger \phi_{D }}+\Gamma_{W^+ W^- \rightarrow \phi_{D }^\dagger \phi_{D }} \nn\\
&&+\Gamma_{Z Z \rightarrow \phi_{D }^\dagger \phi_{D }}+\Gamma_{b \bar{b}  \rightarrow \phi_{D }^\dagger \phi_{D }}+\Gamma_{t \bar{t}  \rightarrow \phi_{D }^\dagger \phi_{D }}\Big].\label{BMsc4} 
\eea
The corresponding yield evolution in the co-moving volume  can be written as, 
\bea
\frac{dY_{\phi_{D }}}{d z}&= &\frac{z^{4}}{sH} \Big[(4-3\theta (z-z_{EW})) \Gamma_{h h \rightarrow \phi_{D }^\dagger \phi_{D }}+\Gamma_{S S  \rightarrow \phi_{D }^\dagger \phi_{D }}+\Gamma_{N N  \rightarrow \phi_{D }^\dagger \phi_{D }}+\Gamma_{S  \rightarrow \phi_{D }^\dagger \phi_{D }}\nn\\
&&+ \theta (z-z_{EW}) \Big[\Gamma_{h  \rightarrow \phi_{D }^\dagger \phi_{D }}+\Gamma_{h S  \rightarrow \phi_{D }^\dagger \phi_{D }}+\Gamma_{W^+ W^- \rightarrow \phi_{D }^\dagger \phi_{D }}\nn \\
&&	 +\Gamma_{Z Z \rightarrow \phi_{D }^\dagger \phi_{D }}+\Gamma_{b \bar{b}  \rightarrow \phi_{D }^\dagger \phi_{D }}+\Gamma_{t \bar{t}  \rightarrow \phi_{D }^\dagger \phi_{D }}\Big]\Big]. \label{yieldsc4}
\eea
\begin{figure}[h]
	\begin{center}
		\mbox{\subfigure[]{\includegraphics[width=0.56\linewidth,angle=-0]{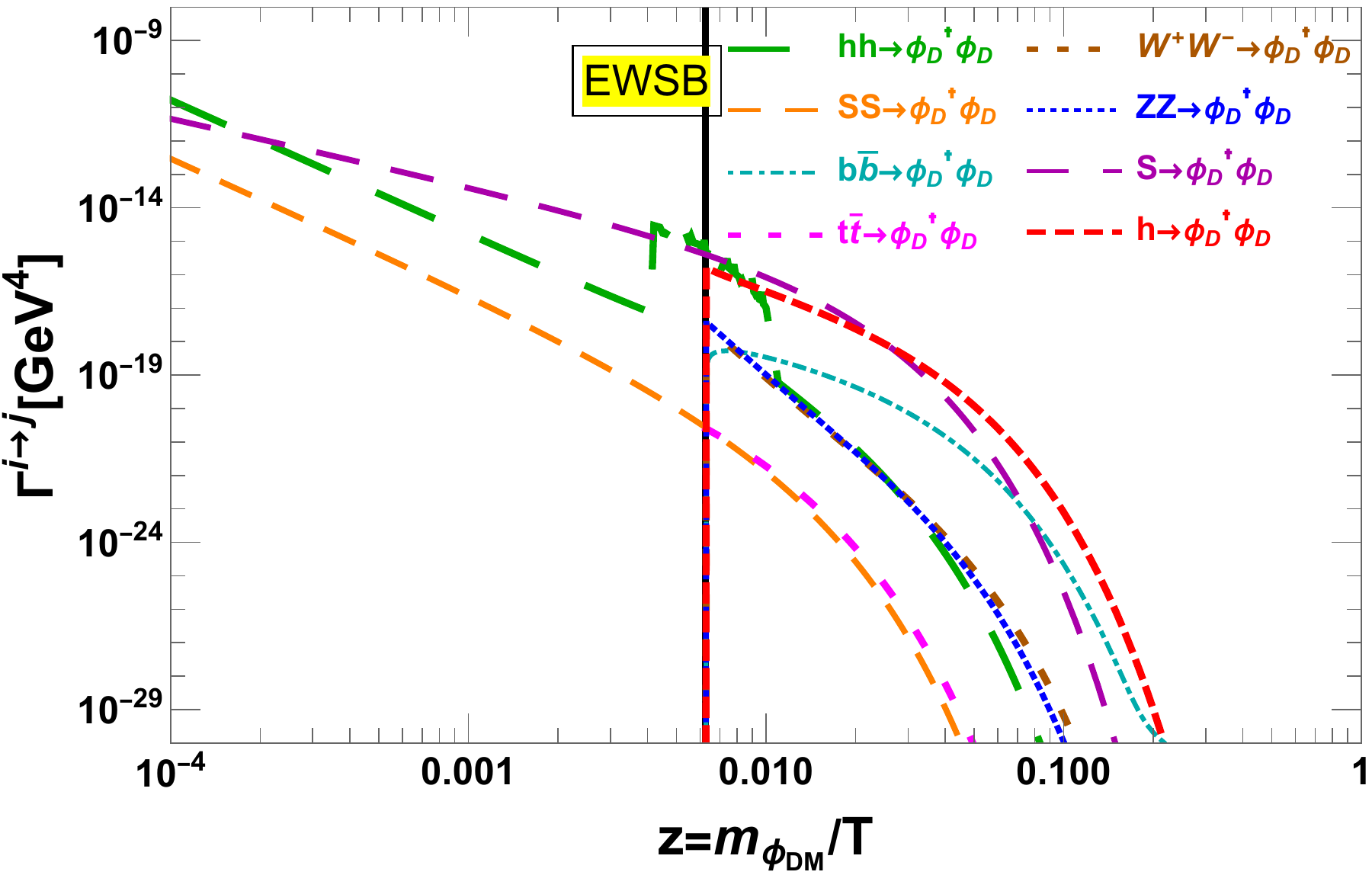}\label{af1}}
			\subfigure[]{\includegraphics[width=0.56\linewidth,angle=-0]{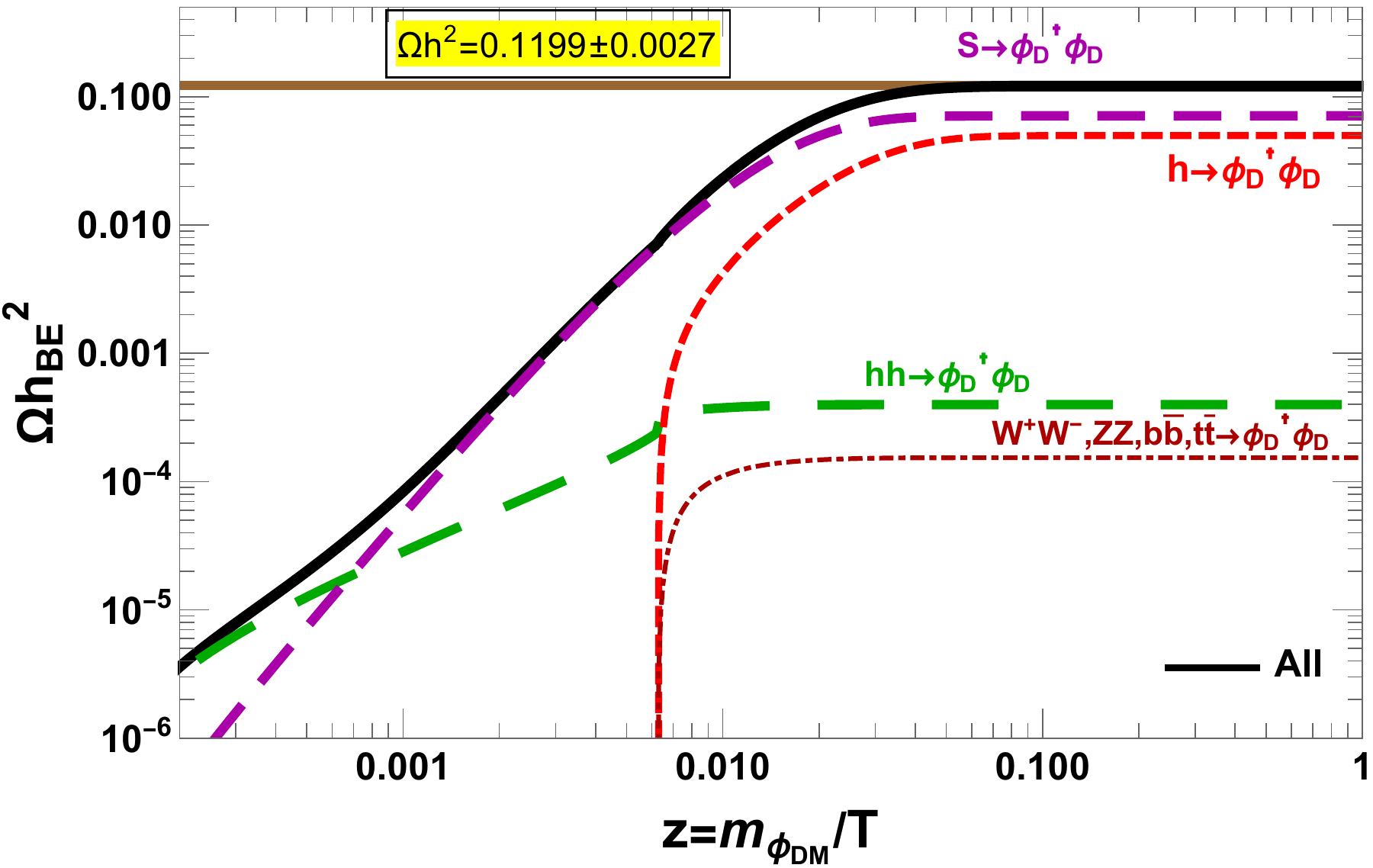}\label{af2}}}
			\mbox{\subfigure[]{\includegraphics[width=0.56\linewidth,angle=-0]{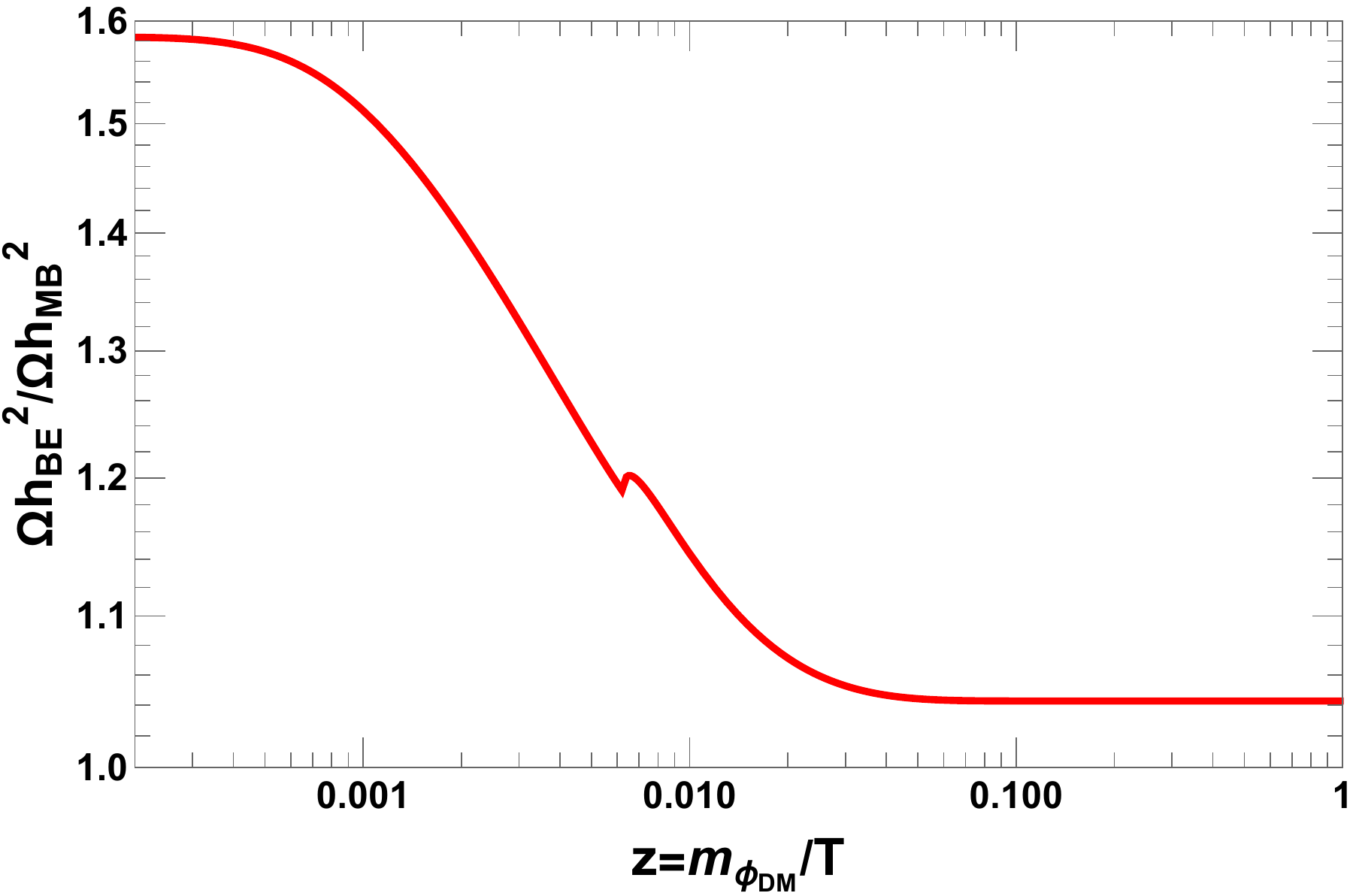}\label{af3}}}
		\caption{The figures correspond to {\it Scenario-4}. Fig.~\ref{af1} shows the relativistic reaction rates for the process $ h h, SS \to \phi_D \phi_D$, and $h, S  \to \phi^*_D \phi_D$. Fig.~\ref{af2} shows the individual contributions to the relic density, and the total  relic density. The brown horizontal line represents the present experimentally measured relic density \cite{Ade:2015xua}. Fig.~\ref{af3} shows the relative enhancement in the relic density with respect to Maxwell-Boltzmann distribution.}\label{Fig6}
	\end{center}
\end{figure}
We include all the contributions as mentioned in Eq.~\eqref{BMsc4} in our numerical analysis. We show the relativistic reaction rates, relic density and the relative enhancement of relic density in Fig.~\ref{Fig6}.  In Fig.~\ref{af1}, we illustrate the relativistic reaction rates for different processes corresponding to the production of $\phi_{D }$.\ Similar to \sn3, here also we observe the resonant enhancement around EWSB for $hh \to \phi^*_{D }\phi_{D }$ annihilation (green bump). The reaction rate for $S \to \phi_{D }^*\phi_{D }$ dominates until almost EWSB, after which $h \to \phi_{D }^*\phi_{D }$ takes over. Similar to the previous scenario, we avoid any over-counting of $S$ on-shell production, by removing it from $h h \to \phi_{D }^* \phi_{D }$ annihilation process.\ Such procedure has also been followed for other similar processes, such as, $b \bar{b} \to \phi_{D }^* \phi_{D }$(mediated by $h$). Since we consider $m_{\phi_{DM}}=1$ GeV, $h \to \phi^*_D \phi_D$ decay (red line) opens up at $z\approx 6.25\times10^{-3}$ during EWSB, when SM Higgs boson takes vev.\newline 
From Fig.~\ref{af2}, one can see that the contribution of the $h,\, S$ decay in relic density are nearly equal.\ These are the dominant production modes.\ Other SM annihilations, such as, \ $b\bar{b},\, t \bar{t},\, W^+ W^{-}, \, ZZ  \rightarrow \phi^*_{D } \phi_{D } $ open up only after EWSB. However, their contributions are much suppressed in this scenario. \newline
In Fig.~\ref{af3}, we show the relative enhancement of the relic abundance using BE and MB distribution. The dark matter production is dominated by SM Higgs boson $h$ annihilation at a very early epoch $z\lesssim 2\times 10^{-4}$ when the ratio is very high around $1.6$. The ratio then lowers down before it saturates at $1.02$. A kink appears in the ratio  at EWSB, {\it i.e.},  $z\approx 6 \times 10^{-3}$, which will be explained in the later on in the subsection. 
\subsection{{\it Scenario-5}}\label{sc5}
\begin{table}[H]
	\centering
	\renewcommand\arraystretch{1.1}
	\begin{tabular}{c| c  c  c | c  c c c c}
		\hline\hline
		\multirow{2}{*}{ \textit{Scenario}  } 	& \multicolumn{3}{c}{Masses in GeV} & \multicolumn{5}{|c}{Couplings} \\
		& $m_S$ & $m_N$ & $m_{\phi_{DM}}$  & $y_N$ & $\lambda_{SD}$ & $\lambda_{Sh}$ & $\lambda_{NS}$& $\lambda_{Dh}$\\ \hline
		\textit{5} & 200  & 300 & 1 & $10^{-7}$ & $3.6\times10^{-13}$ & $ 6\times 10^{-6}$&$0.053$ &$1.24\times10^{-11}$\\
		\hline\hline
	\end{tabular}
	\caption{\MM{The choices of masses and couplings for {\it Scenario-5.}} }\label{Table5}
\end{table}
In this scenario, the primary contribution to relic density arises from the  decay process $h \to \phi^*_D \phi_D$ as we choose higher $\lambda_{Dh}$. Similar to \sn4, here also we choose a light dark matter with mass $m_{\phi_{DM}}=1$ GeV, as shown in Table~\ref{Table5}.   The Boltzmann and the yield equations have the same form as in \sn4, so we follow Eq.~\eqref{BMsc4} and Eq.~\eqref{yieldsc4} for our numerical analysis. 
\begin{figure}[h]
	\begin{center}
		\mbox{\subfigure[]{\includegraphics[width=0.56\linewidth,angle=-0]{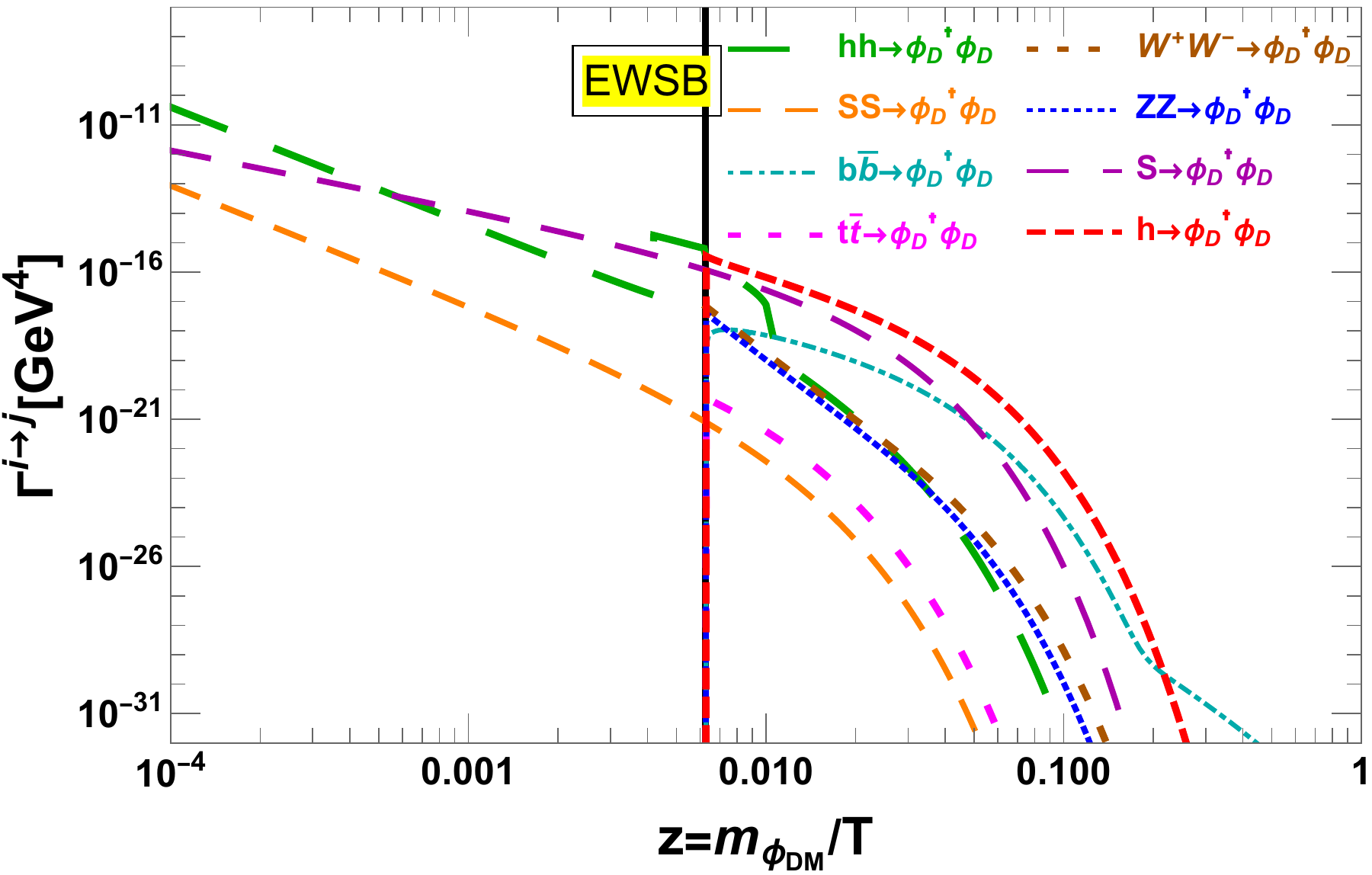}\label{sf1}}
			\subfigure[]{\includegraphics[width=0.56\linewidth,angle=-0]{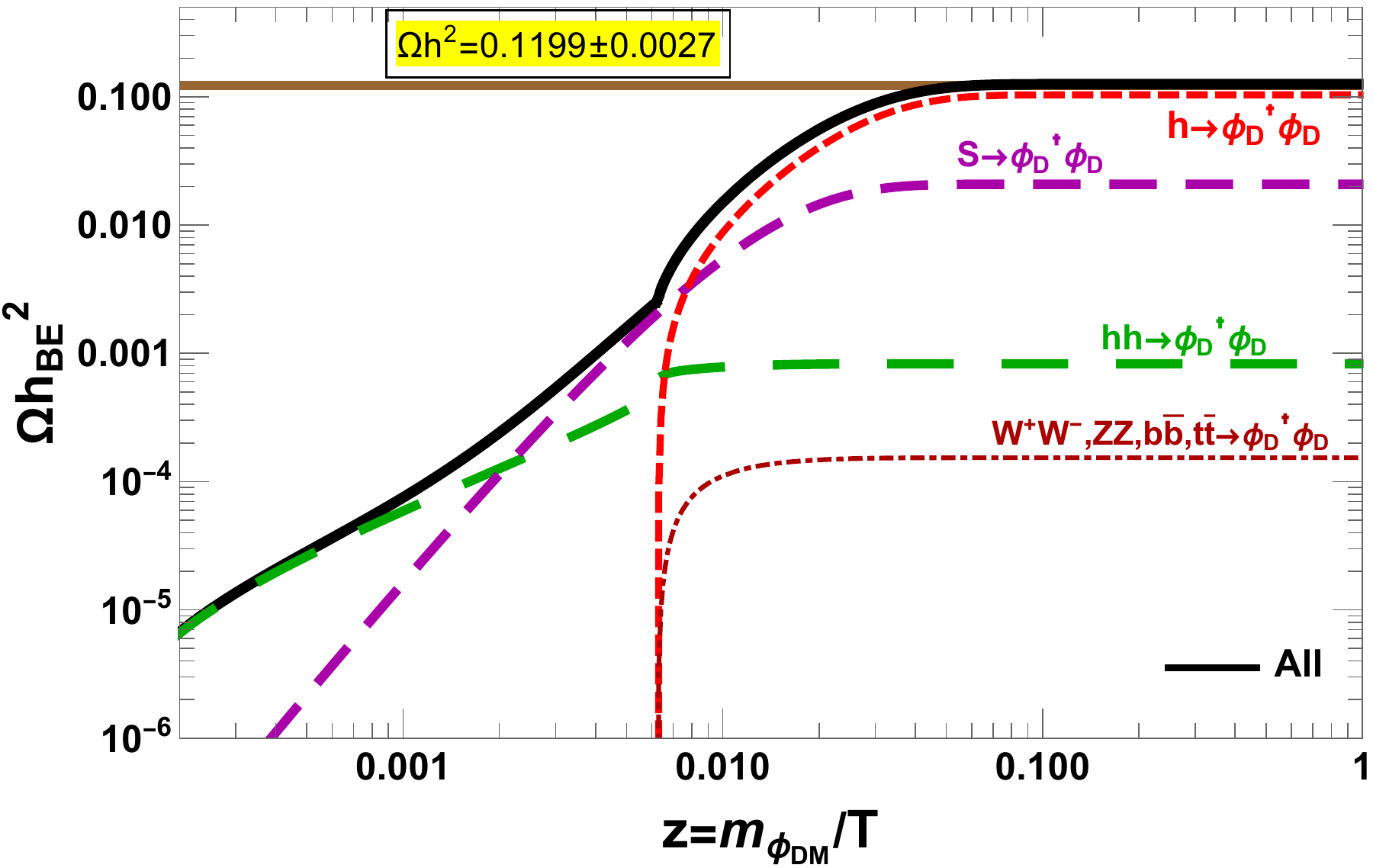}\label{sf2}}}
		\mbox{\subfigure[]{\includegraphics[width=0.56\linewidth,angle=-0]{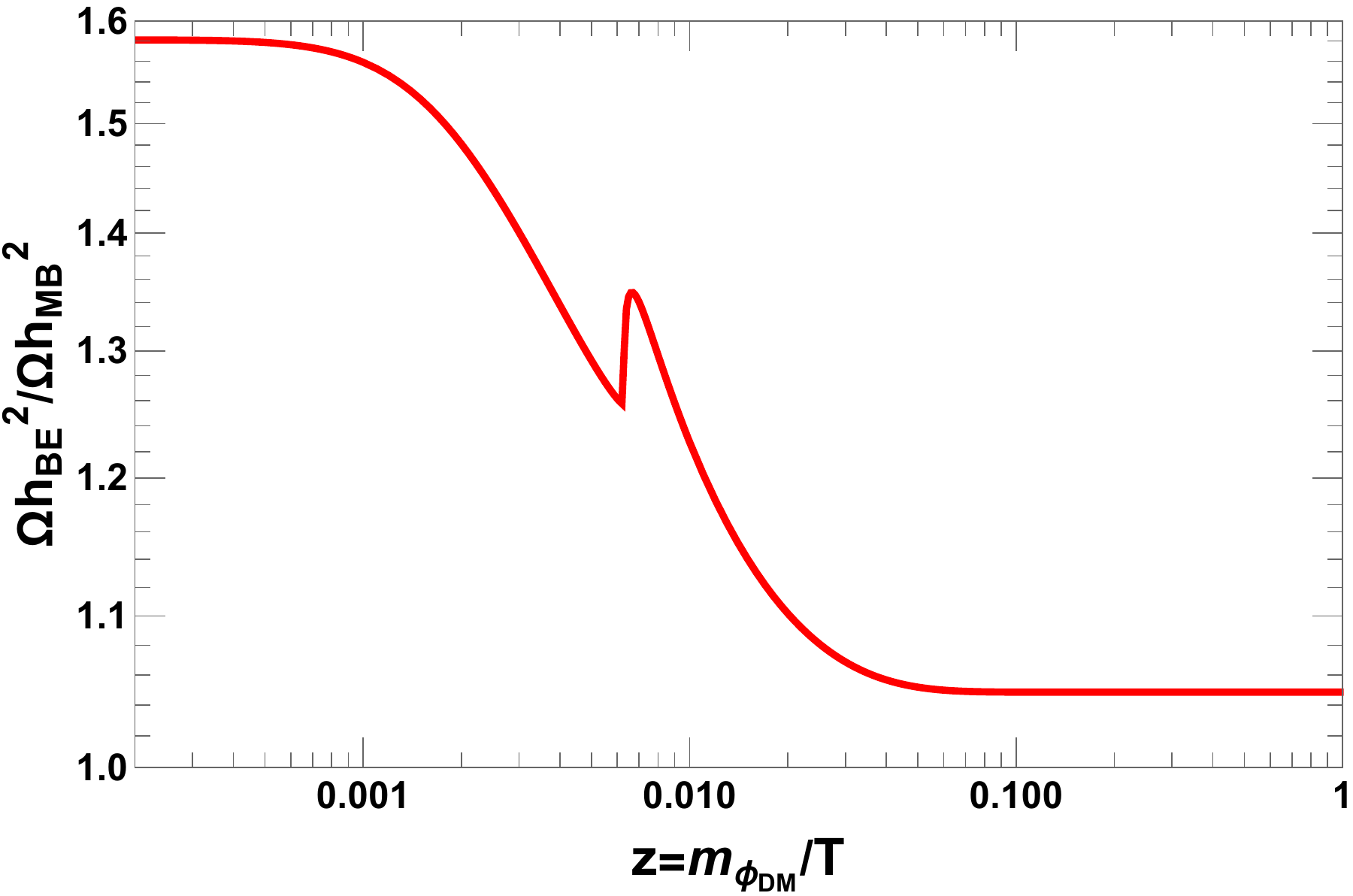}\label{sf3}}
			\subfigure[]{\includegraphics[width=0.56\linewidth,angle=-0]{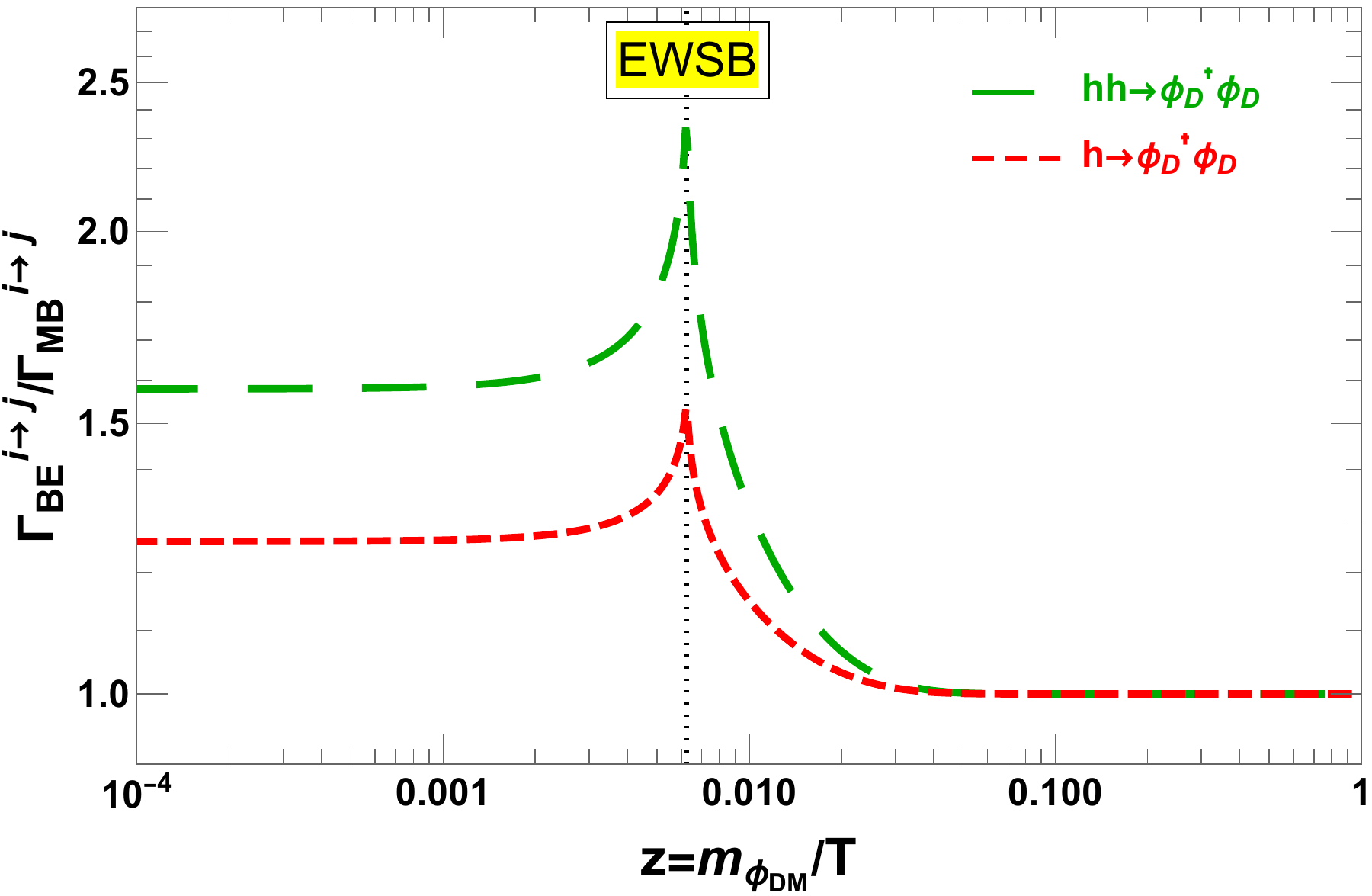}\label{sf4}}}
		\caption{The figures correspond to {\it Scenario-5}. Fig.~\ref{sf1} shows the relativistic reaction rates for the process $ h h, SS \to \phi^*_D \phi_D$, and $h, S  \to \phi_D \phi_D$. Fig.~\ref{sf2} shows the individual contributions to the relic density, and the total  relic density. The brown horizontal line represents the present experimentally measured relic density \cite{Ade:2015xua}. Fig.~\ref{sf3} shows the relative enhancement in the relic density as compared to Maxwell-Boltzmann distribution and Fig.~\ref{sf4} shows the relative enhancement in the respective reaction rates. }\label{Fig7}
	\end{center}
\end{figure}
Fig.~\ref{Fig7}	 shows variation of the different annihilation and decay channels along with the evolution of dark matter relic and BE/MB comparison wih $z$. Fig.~\ref{sf1} shows that the annihilation  $h h \to \phi^*_D \phi_D$ is dominant only at a very early epoch,  then the $S$ decay takes over,  and finally at EWSB {\it i.e.,}  $z\approx 6.25\times10^{-3}$, $h$ decay opens up and becomes the most dominate till   $\phi_{DM}$ freezes-in around $T \sim 20$ GeV. Due to the choice of a  light dark matter, the process $h h  \rightarrow \phi^*_{D } \phi_{D } $ mediated by $S$ encounters $s$-channel resonance during EWSB, which is shown by the  green bump. We follow the same prescription as before, where we omit on-shell contribution from the above mentioned process, and consider only contact term and off-shell contributions in the relic density.  Other SM annihilation processes, such as,  $b\bar{b},\, t \bar{t},\, W^+ W^{-}, \, ZZ  \rightarrow \phi^*_{D } \phi_{D } $ (mediated via $h,S$) open up only after EWSB, however, their rates are relatively  small. Similar to the previous scenario, the $b \bar{b} \to \phi^*_D \phi_D$ also entails a resonance, due to mediation of an on-shell $h$, and we again adopt the same prescription as {\it Scenario-4}. Since the $NN$ contribution is very small, we do not show that in  Fig.~\ref{sf1}, and  Fig.~\ref{sf2}. None of the any other SM annihilation channels contain any resonance.

In Fig.~\ref{sf2} different contributions in obtaining the correct dark matter relic are shown. It is seen that $h \to  \phi^*_D \phi_D$ and $S \to  \phi^*_D \phi_D$ are two dominant modes with 83\% and 17\% 
contributions towards attaining the desired dark matter relic in a freeze-in mechanism at  the temperature of 20 GeV. $h \to  \phi^*_D \phi_D$ is the leading contributor due to a larger $\lambda_{Dh}$ as compared to  $\lambda_{SD}$. The other SM contributions $b\bar{b},\, t \bar{t},\, W^+ W^{-}, \, ZZ  \rightarrow \phi^*_{D } \phi_{D } $ are small $\sim 0.1\%$ only.

Fig.~\ref{sf3} depicts the relative enhancement of relic abundance using BE and MB distributions as a function of $z$. At an early epoch, $h h \to \phi^*_D \phi_D$ primarily dominates the relic density leading to a ratio $\simeq 1.6$.  The ratio falls as temperature decreases and  at EWSB, {\it i.e.,} $z\approx 6 \times 10^{-3}$ a distinct kink appears in the ratio. The kink is more pronounced as compared to {\it Scenario-4}, \MM{due to the presence of SM Higgs decay $h \to \phi^*_D \phi_D$ at EWSB. } This kink is due to the sudden jump in the rates as can be seen from Fig~\ref{sf4}.  This is also to note that for the annihilation $h h \to \phi^*_D \phi_D$(via contact) (green line) kink is more pronounced than the decay process  $h \to \phi^*_D \phi_D$ (red line) at EWSB.  The relative enhancement in the relic density varies from $\sim 1.6$ at a lower value of $z$ to $\sim 1.02$ during the freeze-in temperature  of 16.66 GeV.

The nature of the ratio of the relic densities obtained from BE and MB distributions and the appearance of the kink can be understood in the following way. The $1 \to 2$ rate has the following form: 

\begin{equation}
	\Gamma_{1 \rightarrow 2}=\frac{\Gamma M^3}{2 \pi^2}\int_{1}^{\infty}dt\frac{\sqrt{t^2-1}}{e^{\frac{M }{T}t}-1}.   
	\label{eq:kink1}
\end{equation}
For SM Higgs boson ($h$) decay using BE distribution we get 
\begin{equation}
	\Gamma_{h \rightarrow \phi^*_D \phi_D}^{BE}=\frac{\Gamma m^3_h}{2 \pi^2}\int_{1}^{\infty}dt\frac{\sqrt{t^2-1}}{e^{\frac{m_h }{T}t}-1}   \\
	=\frac{\Gamma m^3_h}{2 \pi^2}\int_{1}^{\infty}dt {\sqrt{t^2-1}} e^{-\frac{m_h }{T}t} {(1-e^{-\frac{m_h }{T}t})}^{-1}.
	\label{eq:kink2}
\end{equation}
Substituting $K_1(z)= z\int_{1}^{\infty} \sqrt{x^2-1} e^{-zx} dx$ in the above equation one realises
\begin{equation}
	\Gamma_{h \rightarrow \phi^*_D \phi_D}^{BE}=\frac{\Gamma m^2_h T}{2 \pi^2} \sum_{n=1}^{\infty}  { \frac{1}{n}} K_1\left(n\frac{m_h}{T}\right). 
	\label{eq:kink33}
\end{equation}
Whereas for MB distribution this rather becomes, 
\begin{equation}
	\Gamma_{h \rightarrow \phi^*_D \phi_D}^{MB}=\frac{\Gamma m^2_h T}{2 \pi^2}{K_1\left(\frac{m_h}{T}\right)}. 
	\label{eq:kink3}
\end{equation}
Now we can compare them as follows
\begin{equation}
	\frac{\Gamma_{h \rightarrow \phi^*_D \phi_D}^{BE}}{\Gamma_{h \rightarrow \phi^*_D \phi_D}^{MB}}=\frac{K_1(\frac{m_h}{T})+0.5 K_1(\frac{2 m_h}{T})+0.33K_1(\frac{3 m_h}{T}).. }{K_1(\frac{m_h}{T})}.
	\label{eq:kink4}
\end{equation}
At EWSB when $T=160$ GeV the Higgs mass becomes $m_h=10$ GeV (see Fig~\ref{fmh}). At that point the next to leading order terms contribute substantially to the $\Gamma_{h \rightarrow \phi^*_D \phi_D}^{BE}$ resulting in a ratio $\frac{\Gamma_{h \rightarrow \phi^*_D \phi_D}^{BE}}{\Gamma_{h \rightarrow \phi^*_D \phi_D}^{MB}}=1.472$. However, this is not true for $S$ decay, since mass of $S$ is considerably large $m_S=200$ GeV, which does not lead to this kind of enhancement at EWSB.

Overall, we find that the thermal mass correction to SM Higgs boson and the freeze-in temperature have a large impact in determining the final enhancement factor.  For {\it Scenario-1,2} the freeze-in occurs around EWSB.\ The $h h \to \phi^*_D \phi_D$ rate is substantially large around EWSB (this holds for all scenarios), when BE distribution is being used.  This leads to a large final enhancement in the ratio of relic density $\Omega h^2_{BE}/\Omega h^2_{MB}$. However, for {\it Scenario-3,4,5} the freeze-in occurs at a later epoch than  EWSB and hence the final enhancement factor is relatively small. This phenomena is rather generic and one can see it from the first term of Eq.~\eqref{eq:kink33} which corresponds to reaction rate using MB distribution (see Eq.~\eqref{eq:kink4}) and other terms in the series provide the correction to the reaction rate. The behaviour of $K_1(z)$ plays significant role in determining the relative enhancement in the reaction rates and hence the relative enhancement in the relic densities. Whenever decaying particle's mass becomes less than temperature  {\it i.e.,} $m_h<<T$, $z<<1$, the correction gives significant contribution to the rates. This occurs in the limit where the Bessel function  {\it i.e.,} $K_1(z)\approx 1/z >>1$. Similarly, for $m_h>>T$  {\it i.e.,} $z>>1$ limit,  the Bessel function $K_1(z)\approx \frac{e^{-z}}{\sqrt{z}} <<1$ \cite{Blennow:2013jba}, which results into the same reaction rate for both BE and MB distributions.
\section{Conclusion \label{conclu}}
	
		We analyse the freeze-in production of a scalar dark matter in an extended gauged $B-L$ model where a complex scalar field $\phi_D$ is the dark matter candidate. To evaluate its relic abundance,  we follow a relativistic \MMc{formalism}, where we consider  \MMc{Bose-Einstein  and Fermi-Dirac  statistics}.  Due to a very tiny charge  of the dark matter under $B-L$ gauge symmetry, that in turn leads to a suppressed interaction  of dark matter with the $B-L$ gauge boson, its production from the $Z_{BL}$ gauge boson  is negligible, and hence not important for our study. We rather focus on the annihilation and decay of the SM and $B-L$ Higgs boson $hh, SS \to \phi^*_D \phi_D$, and $h,S \to \phi^*_D \phi_D$ that contribute primarily to the relic density.   In evaluating the annihilation contribution, we \MMc{consider} all possible processes, namely, the contribution from the four-point contact interaction \MMc{involving Higgs/$B-L$ Higgs boson and dark matter, that directly contribute to $h h/S S \to  \phi^*_D\phi_D$,} as well as, any other $s$-channel mediated processes. \MMc{ The $t$- channel diagrams give suppressed contribution in our case, and hence have not been considered.} Additionally, we also consider the annihilation of SM particles, such as $W^+W^-, ZZ, b \bar{b}, t\bar{t} \to \phi^*_D \phi_D$ that contribute at most by 1$\%$ to the relic density. Depending on the mass of dark matter, and the primary production mechanism, we consider five different scenarios {\it Scenario 1-5}. We show that thermal  correction to  the SM Higgs mass has a significant  impact on different annihilation channels, such as,  $h h \to \phi^*_D \phi_D$,   $b \bar{b} \to \phi^*_D \phi_D$, where, a few of these processes undergo resonance enhancement  in their respective reaction rates, due to the on-shell mediation of $h,S$ states. We consider a fixed mass $m_S=200$ GeV for this study. The entire discussion have been sub-divided into the following few scenarios. 
	
	\begin{itemize}
		\item
		In {\it Scenario-1} and {\it Scenario-2}, we explore the freeze-in production assuming a dark matter with mass $250,150$ GeV, \MMc{respectively}. The primary dark matter production mechanism is the $h h \to \phi^*_D \phi_D$ for the 1st scenario, and $SS \to \phi^*_D \phi_D$ for the second scenario. Due to the choice of the mass of dark matter, neither the SM or $B-L$ Higgs boson decays to dark matter state.  We find for a large $\lambda_{Dh}$ coupling, the $h h \to \phi^*_D \phi_D$ dominates the production, \MMc{whereas}  for \MMc{ a large } $\lambda_{SD}$, the $S S \to \phi^*_D \phi_D$ gives dominant contribution. 	
		\item
		In {\it Scenario-3}, we consider dark matter mass to be  $80$ GeV. This serves as one of the illustrative cases, where both the annihilation of SM Higgs boson $h h \to \phi^*_D \phi_D$, and the decay of $B-L$ Higgs boson $S \to \phi^*_D \phi_D$ contribute to the relic density.  The SM Higgs annihilation channel serves as a primary production channel at an early epoch, while the $S \to \phi^*_D \phi_D$ channel becomes dominant at a later epoch. Due to choice of dark matter mass, the Higgs boson decay is kinematically forbidden. 	
		\item
		In {\it Scenario-4} and {\it Scenario-5}, we consider the dark matter mass to be significantly lower than the SM Higgs boson  mass, $m_{\phi_{DM}}=1$ GeV.  For {\it Scenario-4}, both the $h,S$ decays contribute almost equally to the relic density. For {\it Scenario-5}, the primary production mode is  the  SM Higgs boson  decay to dark matter particle. 
	\end{itemize}

This article presents a comparison between the relic density obtained by using BE statistics, with the one obtained by using MB statistics. We see for the annihilation dominated scenarios, {\it Scenario-1,2}, where  freeze-in occurs during EWSB, the final ratio of relic density obtained using BE and MB statistics is large, $\mathcal{R}=\frac{\Omega_{\rm BE} h^2}{\Omega_{\rm MB} h^2}$ varies  between 1.42-1.62.  For the other three scenarios, where the decay of SM and $B-L$ Higgs bosons dominate the relic density and freeze-in occurs at a much later epoch for the {\it Scenario-3,4,5}, the  enhancement factor is much less $\simeq 1.04$. 
	 
This effect is inherently linked with thermal mass correction of SM Higgs boson, which is considered in this study.  However for the scenarios considered here, the thermal mass correction to $S$ and dark matter are not relevant. We consider the EWSB as a crossover and explore the effect of thermal mass correction of SM Higgs boson on dark matter abundance. It is noticed that due to the low mass of the SM Higgs boson {\it i.e.,} $m_h =10$ GeV at EWSB temperature $T=160$ GeV,  the relativistic reaction  rate for $h h \to \phi^*_D \phi_D$ via contact term becomes significantly enhanced.\ This occurs as the correction terms in the relativistic reaction rate obtained using BE statistics  become significantly large during EWSB. This results in an enhanced reaction rate of $h h \to \phi^*_D \phi_D$ during EWSB, when BE statistics being used in particular relevant for the light dark matter mass. The relative enhancement is more pronounced for annihilation (almost $\simeq 2.3$), as compared to decay ($\simeq 1.5$). 

 The relative enhancement in the reaction rates also result in a distinct kink in the $\mathcal{R}$ around EWSB for \sn{4,5}, where SM Higgs boson decay or annihilation processes contribute significantly  in the dark matter relic abundance.  For {\it Scenario-2,3} since the $S$ decay or annihilation are dominant, therefore, we do not see such an intermediate kink in the ratio of relic density plot. 
 
 We conclude with the observations that  quantum statistics, along with the thermal mass correction are essential to capture these enhancement effects in dark matter relic density in freeze-in scenario which otherwise would be overlooked.
 
 Finally, we make qualitative remarks about few other possible extensions. Allowing a large dark matter self-interaction, its coupling with the SM  and BSM Higgs field, and also a  much suppressed gauge coupling $g_{BL}$ will lead to a different freeze-in dynamics. In this work, we have assumed dark matter self-interaction is negligible. For a large  dark matter self-interaction,  dark matter will thermalise with itself via  $2 \to 4$ processes in the early Universe \cite{ Arcadi:2019oxh,Bernal:2015ova, Bernal:2015xba}. We have also considered, that the quartic interactions of the dark matter with SM and BSM Higgs is negligible. For our assumptions about the dark matter self-interaction, and quartic coupling with the scalar  fields,  dark matter in our scenario is non-thermal. Furthermore, due to small couplings associated with quartic interactions their   impact on the thermal correction of dark matter mass is negligible. Allowing a large quartic coupling, the thermal correction to the dark matter mass will  be sizeable, which needs to be included in the study. For a large dark matter self-interaction, $2\to 4$ processes will also be important.  \\
 Furthermore, for a suppressed $g_{BL}$ coupling,  freeze-in dynamics will be much more involved.  For our chosen benchmark points, which includes a large gauge coupling $g_{BL}$, and the charge of BSM Higgs,  the BSM Higgs $S$  quickly thermalises with the SM particles. Hence, sequential freeze-in \cite{Belanger:2020npe}, i.e.,  production of $S$ and then production of the dark matter from $S$ can not be materialised. However, for a very suppressed  gauge coupling $g_{BL}$ by many orders of magnitude, and  also  suppressed quartic interactions of the BSM Higgs with SM Higgs field,  most of the BSM (BSM Higgs, heavy neutrino $N$, $Z_{BL}$ etc) particles in our model will be non-thermal. Their evolutions in the early Universe will be highly dynamic and coupled which will be determined by solving sets of coupled Boltzmann equations. The detail analysis of these few interesting possibilities is beyond the scope of this paper, and  will be explored in a further study. 
	
	\section*{Acknowledgments}
 MM  acknowledges the support from  {\it Indo-French Centre for the Promotion of Advanced Research} (project no: 6304-2). PB thanks IOP Bhubaneswar for the visit during the first part of the collaboration and ANOMALIES 2020. PB also thanks to SERB CORE Grant CRG/2018/004971 and MATRICS Grant MTR/2020/000668 for the financial support towards the work. AR acknowledges SAMKHYA: High-Performance Computing  Facility  provided  by  the  Institute  of  Physics (IoP), Bhubaneswar. The authors thank Prof.\ Takashi Toma for useful correspondence.

\appendix
\section{Relativistic Rates with the Bose-Einstien distribution and Fermi Dirac distribution function \label{appen2}}
The  dark matter can be produced via annihilation and decay, that may occur  in relativistic regime, {\it i.e.,} when the temperature of  thermal bath exceeds the dark matter mass. The incoming  states for a particular production mode can be boson or fermion. Accordingly, either the  Bose-Einstien or Fermi-Dirac distributions are required in the evaluation of the  reaction rates for the relevant processes. 
The relativistic formalism for reaction rates have been derived in \cite{Lebedev:2019ton, Arcadi:2019oxh}. Here, we briefly summarise the results. 
The reaction rate per unit volume  has the generic expression: 
\begin{equation}\Gamma_{a \rightarrow b} = \int  (\prod_{i \epsilon a}\frac{d^{3}p_{i}}{(2\pi)^{3} 2 E_{i} }f({p_{i}}) )  (\prod_{j \epsilon b}\frac{d^{3}p_{j}}{(2\pi)^{3} 2 E_{j} }(1+ f({p_{j}})) ) |M_{a \rightarrow b}|^{2}(2\pi)^{4} \delta^{4}(p_{a}-p_{b}).  \end{equation}
Here $M_{a \rightarrow b}$ is the transition amplitude and $f(p)$ is the momentum distribution function. In thermal equllibrium, $f(p)$ can be written in a covariant form as
\begin{equation} f(p)=\frac{1}{e^\frac{u.p}{T}\pm 1}, u=(1,0,0,0)^{T} ,\end{equation}
where the upper(lower) sign is for fermionic (bosonic) particles. 
The final states can either be in equilibrium or non-equilibrium with thermal bath. For the final states, not in equilibrium with the thermal bath implies a negligible initial abundance, leading to the  final state enhancement factor $1 +  f(p_{j}) \approx 1$. Similarly,  for the final states which are in equilibrium with thermal bath, one can  neglect the pauli-blocking /stimulated emission effects,  {\it i.e.,}   $1 +  f(p_{j}) \approx 1$.
For $2 \rightarrow 2$ processes,  cross section is defined by
\begin{equation}  \sigma(p_1,p_2)=\frac{1}{4F(p_1,p_2)}\int |M_{2\rightarrow 2}|^{2}(2 \pi)^4 \delta^4( p_1 +p_2 -k_1-k_2)\prod_{i=1}^{2} \frac{d^3 k_i}{(2 \pi)^3  2 E_{k_{i}}},\end{equation}
The reaction rate can be written in terms of cross section which is given by
\begin{equation}\Gamma_{2 \rightarrow 2} = (2 \pi)^{-6}\int  d^{3}p_{1} d^{3}p_{2}f({p_{1}}) f({p_{2}}) \sigma(p_1,p_2) v_{mol} ,\end{equation}
where $v_{mol} $ is the moller velocity of the incoming particle, and is given by, 
\begin{equation}
v_{mol}=\frac{F(p_1,p_2)}{E_1 E_2}=\frac{\sqrt{(p_1.p_2)^2-m_1^2 m_2^2}}{E_1 E_2}.
\end{equation}
The reaction rate can be easily evaluated in centre of mass (CM) frame. See \cite{Lebedev:2019ton, Arcadi:2019oxh} for the details. Following \cite{Lebedev:2019ton, Arcadi:2019oxh},  we define two new variables $p= (p_1 + p_2)/2$  and $k= (p_1 - p_2)/2$ for a pair of momenta $p_1$ and $p_2$.
The vector $p$ can be Lorentz  transformed to the form 
\begin{equation}	
\begin{bmatrix}
E\\
0\\
0\\
0
\end{bmatrix}= \Lambda(p)^{-1} p.	\end{equation}
In the above, $ E $ represents  the particle energy in CM frame.  In terms of half of the centre of mass energy $E$, rapidity $\eta$ and angular coordinates $\theta$,$\phi$, the vector $p$  can be expressed as \cite{Lebedev:2019ton}
\begin{equation}
\begin{split}
p^0&= E\cosh{\eta},\\
p^1&= E\sinh{\eta}\sin{\theta}\sin{\phi},\\
p^2&= E\sinh{\eta}\sin{\theta}\cos{\phi},\\
p^3&= E\sinh{\eta}\cos{\theta}.
\end{split}
\end{equation}
\subsection{Annihilation  \label{annihilation}}
The reaction rate  for $ ab\rightarrow cd$ processes of incoming bosons is given  by
\small\begin{equation}
\begin{aligned}
\Gamma_{2 \rightarrow 2}^{BE} = &\frac{T}{4 \pi^4}\int_{E^{min}_1}^{\infty}dE E^2 \int_{0}^{\infty} d\eta \frac{\sinh\eta}{	e^{\frac{2E \cosh{\eta}}{T}}-1}\ln\left[\frac{\sinh\frac{(E+k_0)\cosh\eta+|k|\sinh\eta}{2 T}}{\sinh\frac{(E+k_0)\cosh\eta-|k|\sinh\eta}{2 T}}\frac{\sinh\frac{(E-k_0)\cosh\eta+|k|\sinh\eta}{2 T}}{\sinh\frac{(E-k_0)\cosh\eta-|k|\sinh\eta}{2 T}}\right]\\ & \times 4 F \sigma^{CM}(E),
\end{aligned}
\label{eq:ab1}
\end{equation}
where $E^{min}_1=max[\frac{m_a+m_b}{2},\frac{m_c+m_d}{2}]$, $|k|=\sqrt{E^2-\frac{m_a^{2}+m_b^{2}}{2}+\frac{(m_a^{2}-m_b^{2})^2}{16 E^2}}$ and $k_{0}=\frac{m_a^{2}-m_b^{2}}{4 E}$.\newline
We derive the reaction rates for  $ 2\rightarrow 2$ processes of  incoming fermions which has the following expression:
\begin{equation}
\begin{aligned}
\Gamma_{2 \rightarrow 2}^{FD} = &\frac{T}{4 \pi^4}\int_{E^{min}_1}^{\infty}dE E^2 \int_{0}^{\infty}d\eta\frac{\sinh\eta}{	e^{\frac{2E \cosh{\eta}}{T}}-1} \ln\left[\frac{\cosh\frac{(E+k_0)\cosh\eta+|k|\sinh\eta}{2 T}}{\cosh\frac{(E+k_0)\cosh\eta-|k|\sinh\eta}{2 T}}\frac{\cosh\frac{(E-k_0)\cosh\eta+|k|\sinh\eta}{2 T}}{\cosh\frac{(E-k_0)\cosh\eta-|k|\sinh\eta}{2 T}}\right] \\ &\times4 F \sigma^{CM}(E). 
\end{aligned}
\label{eq:aaf1}
\end{equation}
At the low temperature  $T$ limit, 
\small$$\begin{aligned}
\ln\left[\frac{\sinh\frac{(E+k_0)\cosh\eta+|k|\sinh\eta}{2 T}}{\sinh\frac{(E+k_0)\cosh\eta-|k|\sinh\eta}{2 T}}\frac{\sinh\frac{(E-k_0)\cosh\eta+|k|\sinh\eta}{2 T}}{\sinh\frac{(E-k_0)\cosh\eta-|k|\sinh\eta}{2 T}}\right]&\approx\frac{2|k|\sinh\eta}{T},\\
\ln\left[\frac{\cosh\frac{(E+k_0)\cosh\eta+|k|\sinh\eta}{2 T}}{\cosh\frac{(E+k_0)\cosh\eta-|k|\sinh\eta}{2 T}}\frac{\cosh\frac{(E-k_0)\cosh\eta+|k|\sinh\eta}{2 T}}{\cosh\frac{(E-k_0)\cosh\eta-|k|\sinh\eta}{2 T}}\right]&\approx\frac{2|k|\sinh\eta}{T}.
\end{aligned}
$$
Therefore, at the low temperature   $T$ limit, Eq.~\eqref{eq:ab1} and Eq.~\eqref{eq:aaf1} reduces to the reaction rate which is equal to the reaction rate obtained by MB distribution,
\small\begin{equation}
\Gamma_{2 \rightarrow 2}^{MB} =\frac{T}{4 \pi^4}\int_{E^{min}_1}^{\infty}dE E|k|K_1\left(\frac{2E}{T}\right) 4 F \sigma^{CM}(E).
\end{equation}
For the incoming particles having same mass $ m=m_a=m_b$ , Eq.~\eqref{eq:ab1} and Eq.~\eqref{eq:aaf1}  reduce to 
\begin{equation}
\begin{aligned}
\Gamma_{2 \rightarrow 2}^{BE} = \frac{T}{4 \pi^4}\int_{E^{min}_2}^{\infty}dE E^2 \int_{0}^{\infty}d\eta\frac{2\sinh\eta}{	e^{\frac{2E \cosh{\eta}}{T}}-1}\ln\left[\frac{\sinh\frac{E\cosh\eta+\sqrt{E^2-m^2}\sinh\eta}{2 T}}{\sinh\frac{E\cosh\eta-\sqrt{E^2-m^2}\sinh\eta}{2 T}}\right] 4 F \sigma^{CM}(E)
\end{aligned}
\label{eq:bb2},
\end{equation}
for the incoming   bosons where $E^{min}_2=max[m,\frac{m_c+m_d}{2}]$ and 
\begin{equation}
\begin{aligned}
\Gamma_{2 \rightarrow 2}^{FD} = \frac{T}{4 \pi^4}\int_{E^{min}_2}^{\infty}dE E^2 \int_{0}^{\infty}d\eta\frac{2\sinh\eta}{	e^{\frac{2E \cosh{\eta}}{T}}-1}\ln\left[\frac{\cosh\frac{E\cosh\eta+\sqrt{E^2-m^2}\sinh\eta}{2 T}}{\cosh\frac{E\cosh\eta-\sqrt{E^2-m^2}\sinh\eta}{2 T}}\right] 4 F \sigma^{CM}(E).
\end{aligned}
\label{eq:bf2}
\end{equation}
for the incoming  fermions.	At low temperature  $T$ limit, the  Eq.~\eqref{eq:bb2} and  Eq.~\eqref{eq:bf2} reduces to the reaction rate which is equal to the reaction rate obtained by MB distribution, 
\small\begin{equation}
\Gamma_{2 \rightarrow 2}^{MB} =\frac{T}{4 \pi^4}\int_{E^{min}_2}^{\infty}dE E\sqrt{E^2-m^2} K_1\left(\frac{2E}{T}\right) 4 F \sigma^{CM}(E).
\end{equation}

\subsection{Fusion}
The reaction rate for $aa\to b$ for incoming fermion is given by
\begin{equation}
\begin{aligned}
\Gamma_{2 \rightarrow 1}^{FD} = \frac{T}{16 \pi^3}\theta(m_b-2 m_a) \int_{0}^{\infty}d\eta\frac{m_b \sinh\eta}{e^{\frac{m_b \cosh{\eta}}{T}}-1}\ln\left[\frac{\cosh\frac{m_b\cosh\eta+\sqrt{m_{b}^2-4 m_{a}^2}\sinh\eta}{4 T}}{\cosh\frac{m_b\cosh\eta-\sqrt{m_{b}^2-2 m_{a}^2}\sinh\eta}{4 T}}\right]|\mathcal{M}_{2\to1}|^2
\end{aligned}
\label{eq:fg1},
\end{equation}
Similary, the reaction rate for $aa\to b$ for incoming boson is given by
\begin{equation}
\begin{aligned}
\Gamma_{2 \rightarrow 1}^{BE} = \frac{T}{16 \pi^3}\theta(m_b-2 m_a) \int_{0}^{\infty}d\eta\frac{m_b \sinh\eta}{e^{\frac{m_b \cosh{\eta}}{T}}-1}\ln\left[\frac{\sinh\frac{m_b\cosh\eta+\sqrt{m_{b}^2-4 m_{a}^2}\sinh\eta}{4 T}}{\sinh\frac{m_b\cosh\eta-\sqrt{m_{b}^2-4 m_{a}^2}\sinh\eta}{4 T}}\right]|\mathcal{M}_{2\to1}|^2
\end{aligned}
\label{eq:fg2}.
\end{equation}
At the low temperature  $T$ limit,
\small$$
\begin{aligned}
\ln\left[\frac{\cosh\frac{m_b\cosh\eta+\sqrt{m_{b}^2-4 m_{a}^2}\sinh\eta}{4 T}}{\cosh\frac{m_b\cosh\eta-\sqrt{m_{b}^2-2 m_{a}^2}\sinh\eta}{4 T}}\right]=
\ln\left[\frac{\sinh\frac{m_b\cosh\eta+\sqrt{m_{b}^2-4 m_{a}^2}\sinh\eta}{4 T}}{\sinh\frac{m_b\cosh\eta-\sqrt{m_{b}^2-4 m_{a}^2}\sinh\eta}{4 T}}\right]\approx\frac{\sqrt{m_{b}^2-4 m_{a}^2}\sinh\eta}{2T}
\end{aligned}.
$$	
Therefore, at the low temperature  $T$ limit, Eq.~\eqref{eq:fg1} and  Eq.~\eqref{eq:fg2} simplifies to a form which is equal to the fusion rate obtained by MB distribution, 
\begin{equation}
\begin{aligned}
\Gamma_{2 \rightarrow 1}^{MB} = \frac{T}{32 \pi^3}\theta(m_b-2 m_a)\sqrt{m_{b}^2-4 m_{a}^2}K_1\left(\frac{m_b}{T}\right) |\mathcal{M}_{2\to1}|^2
\end{aligned}.
\end{equation}

\subsection{Decay}
Considering decay mode,  the reaction rate for this process is given by
\begin{equation}\Gamma_{1 \rightarrow 2} = \int  \frac{d^{3}p_{1}}{(2\pi)^{3} 2 E_{1} }f({p_{1}}) \int  \frac{d^{3}p_{2}}{(2\pi)^{3} 2 E_{2} } \frac{d^{3}p_{3}}{(2\pi)^{3} 2 E_{3} }|M_{1 \rightarrow 2}|^{2}(2\pi)^{4} \delta^{4}(p_{1}-(p_{2}+p_3)) 
\label{eq:qftd}, \end{equation}
The general expression of decay width is given by
\begin{equation}
\Gamma=\frac{1}{2 M}\int\frac{d^{3}p_{2}}{(2\pi)^{3} 2 E_{2} } \frac{d^{3}p_{3}}{(2\pi)^{3} 2 E_{3} }|M_{1 \rightarrow 2}|^{2}(2\pi)^{4} \delta^{4}(p_{1}-(p_{2}+p_3)), 
\end{equation}
So, the  Eq.~\eqref{eq:qftd} can be written as
\begin{equation}\Gamma_{1 \rightarrow 2} = \int  \frac{d^{3}p_{1}}{(2\pi)^{3} 2 E_{1} }f({p_{1}})   2 M \Gamma
\label{eq:qftd2} . \end{equation}
Using the Eq.~\eqref{eq:qftd2}, we can write the decay rate for the  mother particle being  boson having mass $M$  as 
\begin{equation}
\begin{aligned}
\Gamma_{1 \rightarrow 2}^{BE}=\frac{\Gamma M^3}{2 \pi^2}\int_{1}^{\infty}dt\frac{\sqrt{t^2-1}}{e^{\frac{M }{T}t}-1}\  \overset{\frac{M}{T}<<1}{\approx}\ \frac{\Gamma M^2 T}{2 \pi^2} \sum_{n=1}^{\infty} {\frac{1}{n}}K_1\left(n\frac{M}{T}\right)
\end{aligned}
\label{eq:d1},   
\end{equation}
Similarly, for the  mother particle being fermion having mass $\textrm{M}$, the decay rate is given by
\begin{equation}
\Gamma_{1 \rightarrow 2}^{FD}=\frac{\Gamma M^3}{2 \pi^2}\int_{1}^{\infty}dt\frac{\sqrt{t^2-1}}{e^{\frac{M }{T}t}+1}\ 
\overset{\frac{M}{T}<<1}{\approx}\ \frac{\Gamma M^2 T}{2 \pi^2} \sum_{n=1}^{\infty} {\frac{(-1)^{n+1}}{n}}K_1\left(n\frac{M}{T}\right) 
\label{eq:d2}.  
\end{equation}
At the low temperature   $T$,  Eq.~\eqref{eq:d1} and Eq.~\eqref{eq:d2} reduces to a simpler form which is equal to the decay rate obtained by MB distribution,
\begin{equation}
\Gamma_{1 \rightarrow 2}^{MB}=\frac{\Gamma M^2 T}{2 \pi^2}   {  K_1\left(\frac{M}{T}\right)}  
\end{equation}
Next to the leading order terms contribute in  Eq.~\eqref{eq:d1} and Eq.~\eqref{eq:d2} substantially when $M<<T$. Therefore, the  $\Gamma_{1 \rightarrow 2}^{BE}$ is enchanced and $\Gamma_{1 \rightarrow 2}^{FD}$ is suppressed when compared with $\Gamma_{1 \rightarrow 2}^{MB}$.

\section{Analytical Expressions of relevant cross sections and decay widths.\label{appen1}}
We provide the expressions for the relevant cross sections in half of the centre of mass frame and decay widths, that have been used in  the Boltzmann equations.

\subsection{Cross sections for different processes}

Below we consider that the $B-L$ Higgs boson $ \simeq S $, and SM Higgs boson $\simeq h$, and we neglect the mixings between SM and $B-L$ Higgs boson. As considered in the text, the SM and $B-L$ Higgs boson mixing angle $\sin \alpha=10^{-4}-10^{-5}$. 

\begin{enumerate}
	
	\item $\underline{h h\rightarrow \phi_{\mathrm{D}}^\dagger \phi_{\mathrm{D}}}$
	
	\small$$
	\sigma(E)=\frac{1}{64 \pi E^2} \sqrt{\frac{ E^2-m_{\phi_{DM }}^{2}}{ E^2- m_{h}^{2} }}\left|
	\lambda_{Dh}+\frac{v_{BL}^2\lambda_{ Sh}\lambda_{D h}}{(4 E^2-m_S^2)+im_S\Gamma_S}+\theta(T_{ew}-T)\frac{6v^2\lambda_{h}\lambda_{D h}}{(4E^2-m_h^{2})+im_h\Gamma_h}\right|^2
	$$
	
	The first term is from contact term $\lambda_{DH}{\Phi}^\dagger\Phi \phi^\dagger_D \phi_D$. The second and third terms are via mediation of $S$ and $h$ respectively, where third term appears   only after EWSB.
	
	\item $\underline{S S\rightarrow \phi_{\mathrm{D}}^\dagger \phi_{\mathrm{D}}}$

	\small$$
	\sigma(E)=\frac{1}{64 \pi E^2} \sqrt{\frac{ E^2-m_{\phi_{DM }}^{2}}{ E^2- m_{S}^{2} }}\left|
	\lambda_{SD}+\theta(T_{ew}-T)\frac{v^2\lambda_{ Sh}\lambda_{D h}}{(4 E^2-m_S^2)+im_h\Gamma_h}+\frac{6v_{BL}^2\lambda_{S}\lambda_{SD}}{(4E^2-m_S^{2})+im_S\Gamma_S}\right|^2
	$$	
	
	The first term is from contact term $\lambda_{DS}{S}^\dagger S \phi^\dagger_D \phi_D$. The second and third terms are via mediation of $h$ and $S$ respectively, where second term appears   only after EWSB.
	
	\item $\underline{N N\rightarrow \phi_{\mathrm{D}}^\dagger \phi_{\mathrm{D}}}$
	\small$$
	\sigma(E)=\frac{\left(E^2-m_{N}^{2}\right)}{16 \pi s}  \sqrt{\frac{E^2-m_{\phi_{D M}}^{2}}{E^2- m_{N}^{2}}}\frac{v_{BL}^2\lambda_{NS}^{2} \lambda_{S D }^{2}}{\left(4E^2-m_{S}^{2}\right)^{2}+ m_{S}^{2} \Gamma_{S}^{2}}
	$$	
	
	We consider  the $S$ mediated diagram, and ignore the $Z_{BL}$, $h$ mediated processes. For $Z_{BL}$, the $Z_{BL}-\phi_D-\phi_D$ coupling is vanishingly small. Since we consider the SM and $B-L$ Higgs boson mixing to be 
	very small, hence, we ignore the $h$ mediated diagram. 
	
	\item $\underline{h S\rightarrow \phi_{\mathrm{D}}^\dagger \phi_{\mathrm{D}}}$
	\small$$
	\begin{aligned}
	\sigma(E)=&\frac{\theta(T_{ew}-T)}{16 \pi E} \sqrt{\frac{ E^2-m_{\phi_{DM }}^{2}}{(4 E^2-(m_h+m_S)^2)(4 E^2-(m_h-m_S)^2) }}\times\\ & \left|
	\frac{v v_{BL}\lambda_{ Sh}\lambda_{SD}}{(4 E^2- m_{S}^{2})+im_S\Gamma_S}+\frac{v v_{BL}\lambda_{Sh}\lambda_{Dh}}{(4E^2-m_h^{2})+im_h\Gamma_h}\right|^2
	\end{aligned}
	$$	
	
	This process is mediated by $h,S$. The $WW, ZZ, f\bar{f} \to \phi^{\dagger}_D \phi_D$ processes are mediated via only $h$.  Since the SM and $B-L$ Higgs boson  mixing angle is tiny, we neglect the $S$ mediated contribution. 
	
	\item $\underline{W^+ W^-\rightarrow \phi_{\mathrm{D}}^\dagger \phi_{\mathrm{D}}}$
	
	$$
	\sigma(E)=\frac{m_W^{4}\lambda_{D h}^2\theta(T_{ew}-T)}{324 \pi E^2((4E^2-m_h^{2})^2+m_h^2\Gamma_h^2)} \sqrt{\frac{ E^2-m_{\phi_{DM }}^{2}}{ E^2- m_{W}^{2} }}\left(1+\frac{(4E^2-2 m_W^{2})^2}{8 m_W^{4}}\right)^2
	$$
	
	\item $\underline{Z Z\rightarrow \phi_{\mathrm{D}}^\dagger \phi_{\mathrm{D}}}$
	
	$$
	\sigma(E)=\frac{m_Z^{4}\lambda_{D h}^2\theta(T_{ew}-T)}{324 \pi E^2((4E^2-m_h^{2})^2+m_h^2\Gamma_h^2)} \sqrt{\frac{ E^2-m_{\phi_{DM }}^{2}}{ E^2- m_{Z}^{2} }}\left(1+\frac{(4E^2-2 m_Z^{2})^2}{8 m_W^{4}}\right)^2
	$$
	
	\item $\underline{f \bar{f}\rightarrow \phi_{\mathrm{D}}^\dagger \phi_{\mathrm{D}}}$
	$$
	\sigma(E)=\frac{\left(E^2-m_{f}^{2}\right)\theta(T_{ew}-T)}{32 \pi E^2 n_c} \sqrt{\frac{E^2-m_{\phi_{DM}}^{2}}{E^2- m_{f}^{2}}}\frac{m_f^{2} \lambda_{Dh }^{2}}{\left((4E^2-m_{h}^{2})^2+m_h^2\Gamma_h^2\right)}$$\\
	In the above,  $n_c$ is the color charge, and is 1 for leptons and 3 for quarks.
	
\end{enumerate}

\subsection{Decay widths of $S$}
The expressions for the decay widths of $S$:  
\begin{itemize}
	\item $\Gamma(S \rightarrow \phi^{\dagger}_{D }\phi_{D}) =\frac{\lambda_{SD}^{2} v_{BL}^{2} }{32 \pi m_S}\sqrt{1-\frac{4 m_{\phi_{DM }}^{2}}{m_S^{2}}}$
	
	\item $ \Gamma(S \rightarrow ZZ) =\frac{ m_{S}^{3}}{32 \pi v^2} \sqrt{1-\frac{4 m_{Z}^{2}}{m_{S}^{2}}}\left(1-\frac{4 m_{Z}^{2}}{m_{S}^{2}}+\frac{12 m_{Z}^{4}}{m_{S}^{4}}\right)\sin^{2}{\alpha}$
	
	\item $ \Gamma(S \rightarrow W^{+}W^{-})= \frac{ m_{S}^{3}}{16 \pi v^2}  \sqrt{1-\frac{4 m_{W}^{2}}{m_{S}^{2}}}\left(1-\frac{4 m_{W}^{2}}{m_{S}^{2}}+\frac{12 m_{W}^{4}}{m_{S}^{4}}\right)\sin^{2}{\alpha}$
	
	\item $ \Gamma(S \rightarrow h h )= \frac{\lambda_{Sh}^{2} v_{BL}^{2}}{32 \pi m_S}\sqrt{1-\frac{4 m_{h}^{2}}{m_S^{2}}}$
	
	\item $ \Gamma(S \rightarrow t \bar{t})=\frac{3 m_S m_{t}^2}{8 \pi v^2}\left(1-\frac{4 m_{t}^{2}}{m_S^{2}}\right)^{\frac{3}{2}}\sin^{2}{\alpha} $
	
	\item $ \Gamma(S \rightarrow b \bar{b})= \frac{3 m_S m_{b}^2}{8 \pi v^2}\left(1-\frac{4 m_{b}^{2}}{m_S^{2}}\right)^{\frac{3}{2}}\sin^{2}{\alpha} $
	
	\item $ \Gamma(S \rightarrow NN) =\frac{m_{N}^{2}m_{S}}{16 \pi v_{BL}^2}\left(1-\frac{4 m_{N}^{2}}{m_{S}^{2}}\right)^{\frac{3}{2}}\cos^{2}{\alpha}$
	
	\item $ \Gamma(S \rightarrow N \nu) =\frac{y_{N}^{2} m_{S }}{64 \pi} \left(1-\frac{M_{N}^{2}}{M_{S}^{2}}\right)^{2}\sin^{2}{\alpha}$
\end{itemize}
\subsection{Decay widths of $Z_{BL}$}	
\begin{itemize}
	\item $\Gamma(Z_{BL} \rightarrow \phi_{D }\phi_{D }^\dagger)=\frac{g_{BL}^2 q_{DM}^2 m_{Z_{BL}}}{48 \pi}\left(1-\frac{4m_{\phi_{DM }}^2}{m_{Z_{BL}^2}}\right)^\frac{3}{2}$
	
	\item $\Gamma(Z_{BL} \rightarrow N\bar{N})=\frac{g_{BL}^2 m_{Z_{BL}}}{24 \pi}\left(1-\frac{4m_{N}^2}{m_{Z_{BL}^2}}\right)^\frac{3}{2}$
	
	\item $\Gamma(Z_{BL} \rightarrow f\bar{f})=\frac{n_c g_{BL}^2 m_{Z_{BL}}}{12 \pi}\left(1+\frac{m_{f}^2}{m_{Z_{BL}^2}}\right)\left(1-\frac{4m_{f}^2}{m_{Z_{BL}^2}}\right)^\frac{1}{2}$\\
\end{itemize}

In the above,  $n_c$ represents  the color charge, and is 1 for leptons, and 3 for quarks.

\section{Thermal Correction to SM Higgs Mass}

In this work, we have considered the electroweak phase transition to be crossover in which Higgs remains massive at critical temperature ($T_c=160\, \textrm{GeV}$).\ We have assumed $m_h(T_c)\approx 10\, \textrm{GeV}$.\ When the temperature is greater than the critical temperature $T_c$. In this regime, the mass of Higgs bosons is given by \cite{DeRomeri:2020wng},
\begin{equation}
m_{h}^{2}(T)= c(T^2-T_{c}^2 ) + m_{h}^{2}(T_c).
\end{equation}
For temperature smaller than the critical temperature  $T_c$, the mass of Higgs boson is given by
\begin{equation}
m_{h}^{2}(T)=2 c(T_{c}^2-T^2 ) + m_{h}^{2}(T_c).
\end{equation}
where $c$ is a constant determined by the requirement  $m_h(0)=125.5\, \textrm{ GeV}$. \\

\bibliographystyle{utphys}
\bibliography{bibitem}

\end{document}